\documentstyle[aps,prb,eqsecnum,amsmath,amssymb,bbm,theorem,graphicx,subfigure,citesort,overcite,multicol]{revtex}

\theoremstyle{break}
\newtheorem{theorem}{Theorem}[section]
\newtheorem{assumption}[theorem]{Assumption}

\begin{document}
\draft
\title{A Study of the Antiferromagnetic Phase in the Hubbard Model\\
by means of the Composite Operator Method}
\author{Adolfo Avella\cite{Eavella} and Ferdinando Mancini}
\address{Dipartimento di Fisica ``E.R. Caianiello'' -- Unit\`a INFM di Salerno,\\
Universit\`a degli Studi di Salerno, 84081 Baronissi (SA), Italy}
\author{Roland M\"unzner}
\address{Institut f\"ur Theoretische Physik, Universit\"at
T\"ubingen,\\ Auf der Morgenstelle 14, 72~076 T\"ubingen, Germany}
\date{January 15, 2000}
\maketitle
\begin{abstract} \widetext
We have investigated the antiferromagnetic phase of the 2D, the 3D and the
extended Hubbard models on a bipartite cubic lattice by means of the Composite
Operator Method within a two-pole approximation. This approach yields a fully
self-consistent treatment of the antiferromagnetic state that respects the
symmetry properties of both the model and the algebra. The complete phase
diagram, as regards the antiferromagnetic and the paramagnetic phases, has been
drawn. We firstly reported, within a pole approximation, three kinds of
transitions at half-filling: Mott-Hubbard, Mott-Heisenberg and Heisenberg. We
have also found a metal-insulator transition, driven by doping, within the
antiferromagnetic phase. This latter is restricted to a very small region near
half filling and has, in contrast to what has been found by similar
approaches, a finite critical Coulomb interaction as lower bound at half
filling. Finally, it is worth noting that our antiferromagnetic gap has two
independent components: one due to the antiferromagnetic correlations and
another coming from the Mott-Hubbard mechanism.
\end{abstract}

\begin{multicols}{2} \narrowtext

\section{Introduction}
\label{sec:1}

Since almost half of a century the Hubbard model\cite{HubAnd} is one of the
fundamental models in the theoretical description of strongly correlated
electron systems. Despite its simplicity it has always been taken as one of
the prototypes when modeling electrons in narrow bands. Within this context it
had wide applications in the theory of magnetism
\cite{Nagaoka:66,Tasaki:98,Tasaki:98a} and metal-insulator transitions
\cite{Gebhard:97}. The discovery of the high-$T_c$ superconducting cuprate
materials \cite{Bednorz:86}, where the special properties are to a large
extent due to strong correlations \cite{Anderson:87}, gave a new impetus on the
investigation of the Hubbard model and its derivates as the
$t$--$t^{\prime}$--$U$, the extended Hubbard and the $t$-$J$ models. To gain a
better theoretical understanding of the high-$T_c$ materials, not only the
superconducting phase of the proposed models is studied, but also their normal
phase and such with further broken symmetries. The antiferromagnetic state is
of particular interest, as it is believed that the pairing mechanism in the
cuprates is mainly due to magnetic correlations in an itinerant electronic
system. In addition, the high-$T_c$ compounds show a wide range of
metal-insulator transitions; in particular, a phase transition from an
antiferromagnetic insulator to a superconductor is observed. The
antiferromagnetic phase of the Hubbard model has been subject of intensive
study by both numerical and analytical methods.

Despite the simplicity of the model there are no exact solutions known except
for a few special cases \cite{Lieb:95a,Tasaki:98} and one must recur to
approximate solutions. The \emph{Dynamical Mean Field Theory} (\emph{DMFT})
\cite{Georges:96} and a variety of projection techniques
\cite{Roth:69,Kalashnikov:73,Nolting:72,Fedro:92,Ishihara:94,Mancini:95,Fulde:95,Mori,Zwanzig:61}
are among the most popular approaches to the Hubbard model and its derivates.

The \emph{Composite Operator Method} (\emph{COM})
\cite{Ishihara:94,Mancini:95,Matsumoto} that we present here belongs to the
above mentioned class of projection techniques. Choosing as elements of the
fundamental spinor a set of field operators built up from the electronic ones
(hence the name), the equation of motion for the Green's function of this
spinor are obtained. These equations have been solved by means of both a pole
approach \cite{Ishihara:94,Mancini:95} and a two-site resolvent method
\cite{Matsumoto}. The use of composite operators asks for a careful choice of
the Fock space where the Green's function is realized. We have shown that the
Pauli principle plays a fundamental role in unambiguously fixing the
representation by determining the parameters that appear in the scheme owing
to the non-canonical algebra of the composite operators
\cite{Mancini:00,Mancini:98}. Let us note that by Pauli principle we mean all
relations among operators dictated by the algebra. A comprehensive treatment
of the paramagnetic phase of the Hubbard model within the two-pole
approximation has led to a good agreement with both numerical studies and some
experimental properties of the cuprates \cite{Mancini:95,Mancini}. In the case
of the 1D Hubbard model we have shown that the two-pole approximation
reproduces almost exactly the Bethe ansatz results for the ground state energy
\cite{Avella:98e}.

In the last years an intensive study by the \emph{Spectral Density Approach}
(\emph{SDA}) \cite{Kalashnikov:73,Nolting:72} of the magnetic phases of the
Hubbard model on different lattice types and by varying dimensionality has
been done \cite{Nolting:89,Kellen:90,Herrmann:97a} and the complete magnetic
phase diagram of the Hubbard model on a 3D fcc-lattice was derived
\cite{Herrmann:97a}. The \emph{SDA} corresponds to an expansion of the Green's
function in terms of the electronic spectral moments. It has been shown that
this approach corresponds to the \emph{COM} when a specific choice of the
basic spinor is made and the pole approximation is used \cite{Mancini:98b}.
However, this correspondence is only relative to the functional dependence of
the Green's function and huge differences arise when the representation is not
properly fixed \cite{Avella:98,Mancini:98,Mancini:00}.

To complement the analysis of the paramagnetic phase for the Hubbard model in
the two-pole approximation by the \emph{COM}, we here present a study of the
antiferromagnetic phase within this framework. As the most simple case of
antiferromagnetism, where in general the spin and the translational invariance
are broken, we consider an antiferromagnetic state on a bipartite lattice
characterized by its staggered magnetization. It turns out that in this phase
the simple Hubbard model does not provide enough internal parameters to fix
the representation. We therefore investigate the $t$--$t^{\prime}$--$U$ model
in the limit of $t^{\prime}\rightarrow0$ in two and three dimensions and the
extended Hubbard model in two dimensions.

In the first part of the paper, we develop the analytical background needed
for the calculations. We end up with a closed set of self-consistent equations
for the antiferromagnetic thermal equilibrium state, which respects the Pauli
principle and the particle-hole symmetry. The numerical evaluation of these
self-consistent equations is presented in the second part of the paper. The
complete phase diagrams considering the paramagnetic and the antiferromagnetic
phase for both models and in different dimensionality are calculated, as well
as the density of states and the quasi-particle dispersions. The main results
of the paper can be considered: the presence of three kinds of transitions
(Mott-Hubbard, Mott-Heisenberg and Heisenberg) at half-filling in the plane
$T$-$U$; the existence of two components in the antiferromagnetic gap (one due
to the antiferromagnetic correlations and another coming from the Mott-Hubbard
mechanism); a finite critical value of the Coulomb interaction for the
Mott-Hubbard and Mott-Heisenberg transitions; the strong decay of the N{\'e}el
temperature with doping; a metal-insulator transition away from half-filling
inside the antiferromagnetic phase.

\section{The Hubbard Model within the Two Pole Approximation}
\label{sec:2}

The Hamiltonian for the single-band Hubbard model \cite{HubAnd} with chemical
potential $\mu$ is defined by
\begin{equation}
\begin{split}
H=&\sum_{ij;\sigma }\left( t_{ij}-\mu \delta _{ij}\right) c_\sigma
^{\dagger }\left( i\right) c_\sigma \left( j\right)\\
&\qquad + U\sum_in_{\uparrow }\left( i\right) n_{\downarrow }\left( i\right)
\end{split}
\label{eq:2.1}
\end{equation}
with $\sigma \in \left\{ \uparrow ,\downarrow \right\}$ and where the sum over
the site indices $i,j$ runs over the whole chemical lattice $\{ {\mathbf
R}_{i}\}$, which we consider to be of simple cubic type with lattice constant
$a$. $U$ is the on-site Coulomb interaction. The kinetic part of the
Hamiltonian is given by the nearest neighbor hopping. In $d$ dimensions the
hopping matrix takes the form $t_{ij}=-2\,d\,t\,\alpha_{ij}=-2\,d\,t\frac
1N\sum_{\mathbf{k}}e^{i\mathbf{k}\cdot \left( \mathbf{R}_i-\mathbf{R}_j
\right)} \alpha \left( \mathbf{k}\right)$, with $\alpha \left(
\mathbf{k}\right) =\frac 1d \sum_{l=1}^{d} \cos\left(k_{l}a\right)$.

In the framework of the \emph{COM} we take as basic fields the Hubbard
operators: $\xi _\sigma \left( i\right) = c_\sigma \left( i\right) \left[
1-n_{\overline{\sigma }}\left( i\right) \right]$ and $\eta _\sigma
\left(i\right) = c_\sigma \left( i\right) n_{\overline{\sigma }}\left(
i\right)$, where $n_\sigma \left( i\right) =c_\sigma ^{\dagger }\left(
i\right) c_\sigma \left( i\right)$, describing the basic excitations $n(i)=0
\leftrightarrow n(i)=1$ and $n(i)=1 \leftrightarrow n(i)=2$ on the lattice
site $i$, which are responsible for the leading contributions in the
electronic density of states \cite{Dagotto:94}. Let use define the following
four-component composite field at lattice site $i$:
\begin{equation}
\label{eq:2.2} \Psi\left( i,t\right) =\left(
\begin{tabular}{l}
$\xi _{\uparrow }\left( i,t\right) $ \\
$\eta _{\uparrow }\left( i,t\right) $ \\
$\xi _{\downarrow }\left( i,t\right) $ \\
$\eta _{\downarrow }\left( i,t\right) $
\end{tabular}
\right) \, ,
\end{equation}
where the variable $t$ denotes the time translate of the operator under the
Heisenberg dynamics generated by $H$. The Heisenberg equation of motion for
this basic field $\Psi$ may be split into a part that is linear in the basic
field and the remaining non-linear part
\begin{equation}
\label{eq:2.4} i\frac \partial {\partial t}\Psi\left( i,t\right) = J \left(
i,t\right) =\sum_j\varepsilon \left( i,j\right) \Psi \left( j,t\right) +\delta
J \left( i,t\right)
\end{equation}
by projecting it on the basic field. Then, the matrix $\varepsilon \left(
i,j\right)$ is fixed by
\begin{equation}
\label{eq:2.5}
\begin{split}
\varepsilon \left( i,j\right) =& \sum_l\bigl\langle \bigl\{ J\left(
i,t\right), \Psi^{\dagger }\left(l,t\right)\bigr\}
\bigr\rangle\\
&\times \bigl\langle \bigl\{\Psi\left( l,t\right), \Psi^{\dagger }\left(
j,t\right)\bigr\} \bigr\rangle^{-1} \, ,
\end{split}
\end{equation}
where $\langle\cdot\rangle$ denotes the thermal average on the grand-canonical
ensemble.

To simplify the notational effort we introduce the so-called normalization
matrix $I$ and the so-called $m$-matrix
\begin{eqnarray}
I\left( i,j\right)  &=&\left\langle \left\{ \Psi \left( i,t\right) ,
\Psi^{\dagger }\left( j,t\right) \right\} \right\rangle
\label{eq:2.6}\\
m\left( i,j\right)  &=&\left\langle \left\{ J\left( i,t\right) , \Psi^{\dagger
}\left( j,t\right) \right\} \right\rangle \label{eq:2.7}
\end{eqnarray}
and get the shorthand for the energy matrix $\varepsilon =mI^{-1}$ from
Eq.~\eqref{eq:2.5}.

The pole approximation consists in neglecting the nonlinear part $\delta
J\left( i,t\right)$ in the equation of motion. This is equivalent to the
suppression of the incoherent part in the retarded Green's function matrix
\begin{equation}
\label{eq:2.9}
\begin{split}
S\left( i,j,t\right) = & \theta \left( t\right) \left\langle \left\{
\Psi\left(i, t\right) ,\Psi^{\dagger}\left(j,
0\right)\right\} \right\rangle\\
= & \frac i{2\pi }\Bigl( \frac a{2\pi }\Bigr) ^{d}\int {\rm d}\omega \,
e^{-i\omega t}\int_{\Omega_B}{\rm d}^dk \,e^{i{\mathbf
k\cdot }\left( {\mathbf R}_i-{\mathbf R}_j\right)}\\
& \quad\qquad\times\int_{\Omega _B}{\rm d}^dp \, e^{i{\mathbf p\cdot R}
_i}S\left( {\mathbf k},{\mathbf p},\omega \right) \, ,
\end{split}
\end{equation}
which then satisfies the linearized equation of motion
\begin{equation}
\label{eq:2.10} i\frac \partial {\partial t}S\left( i,j,t\right) = i\delta
\left( t\right) I\left( i,j\right) +\sum_l\varepsilon \left( i,l\right)
S\left( l,j,t\right) \, .
\end{equation}
$d$ is the dimension of the system, $a$ the lattice spacing and $\Omega _B$
the Brillouin zone. The resulting set of algebraic equations for the Fourier
transform of the retarded Green's function may then be solved. The knowledge
of the retarded Green' function will allow us to calculate the correlation
functions $\left\langle {\mathbf \Psi}\left( i,t\right) {\mathbf
\Psi}^{\dagger }\left( j,0\right) \right\rangle$ by means of the spectral
theorem. As already mentioned, the above scheme of truncating the equation of
motion by a projection technique is rather common in the treatment of systems
with strong correlations. When we use as basic fields, as we do in this work,
the operators coming from the hierarchy of the equations of motion
\cite{Avella:98,Mancini:98b}, the \emph{COM} is similar to the \emph{SDA}
\cite{Kalashnikov:73,Nolting:89,Kellen:90,Herrmann:97a}. However, the
\emph{COM} gives the possibility of choosing the basic fields according to the
physics of the system under investigation. This freedom allows a better
control on the dynamical information still present in the generalized
mean-field approximation. Let us emphasize that even in the pole approximation
the electronic self energy is not trivial as in the standard mean field
approach, but a pole expansion of the exact electronic self energy is used.
Another fundamental difference between the \emph{COM} and the approaches
mentioned above consists in the treatment of the higher-order correlation
functions occurring in the energy matrix~\eqref{eq:2.5}, which are not
directly related to elements of the Green's function. As shown in a recent
publication\cite{Mancini:00}, there is no freedom in choosing the equations to
compute those parameters as they have to be used to fix the representation
according to the following relation
\begin{equation}
\label{Paulic} \lim\limits_{
\begin{subarray}{c}
j\rightarrow i \\
t \rightarrow 0^+
\end{subarray}
}S(i,j,t) = \left\langle\Psi(i)\Psi^\dagger(i)\right\rangle \, ,
\end{equation}
where the l.h.s.\ comes from Eq.~\eqref{eq:2.9} and the r.h.s. derives from
the basic algebraic properties of the electronic field algebra -- namely the
\emph{Pauli principle}. Any other choice of the self-consistent equations that
fix those parameters will lead to a representation for the Green's function
where the main symmetries of the system are violated \cite{Avella:98}. In the
case of the Hubbard model, Eq.~\eqref{Paulic} leads to the following
self-consistent equations
\begin{subequations}
\label{eq:2.14}
\begin{align}
\bigl\langle \xi_\sigma (i) \, \eta_\sigma^\dagger (i) \bigr\rangle &= 0
\label{eq:2.14a}\\
\bigl\langle \xi_\uparrow (i) \, \xi_\uparrow^\dagger (i) \bigr\rangle &=
\bigl\langle \xi_\downarrow (i) \, \xi_\downarrow^\dagger (i) \bigr\rangle \,
. \label{eq:2.14b}
\end{align}
\end{subequations}
The use of these equations, in the paramagnetic phase, led to a good agreement
of the \emph{COM} with the numerical results for the local, integrated and
thermodynamic quantities \cite{Mancini:95,Mancini,Avella:98e}.

In contrast to the paramagnetic solution -- characterized by a complete
translational and spin rotational symmetry --, where the number of parameters
in the energy matrix emerging from higher-order correlation functions equals
the number of constrains given by the Pauli principle \eqref{eq:2.14}, an
antiferromagnetic solution with two composite fields, which is characterized
by a broken translational and spin rotational symmetry (they are, however, not
broken in a completely independent way), requires additional parameters to
satisfy all constrains emerging from the algebraic properties of the basic
field operators $\xi$ and $\eta$. It turns out that for the extensions of the
simple Hubbard model \eqref{eq:2.1} by a next nearest neighbor hopping term,
the $t$--$t^{\prime}$--$U$ model, or by a nearest neighbor Coulomb repulsion,
the so-called extended Hubbard model, the number of those parameters equals
the number of `algebraic' constrains in the case of an antiferromagnetic
solution. According to this, in the following we will investigate the
antiferromagnetic solution of the 2D and 3D $t$--$t^{\prime}$--$U$ model in the
limit of $t^{\prime}$ going to zero and of the 2D extended Hubbard model.

For the antiferromagnetic solution the two poles corresponding to the two
composite fields $\xi$ and $\eta$, will split up into four poles due to the
broken translational and spin rotational symmetry.

\section{The Antiferromagnetic Solution for the Hubbard Model} \label{sec:3}

\subsection{The Antiferromagnetic Solution on a Bipartite Lattice}
\label{sec:3.1}

As stated in the above we will study the antiferromagnetic solution of the
Hubbard model by considering the $t^{\prime}\rightarrow0$ limit of the
$t$--$t^{\prime}$--$U$ model. The Hamiltonian of the latter reads as
\begin{equation}
\label{eq:3.1}
\begin{split}
H^{\text{\em tt'U}}= & \sum_{ij;\sigma }\bigl( t_{ij}+t_{ij}^{\prime}-\mu
\delta _{ij}\bigr) c_\sigma ^{\dagger }\left( i\right) c_\sigma \left
( j\right)\\
& \qquad + U\sum_in_{\uparrow }\left( i\right) n_{\downarrow }\left ( i\right)
\, .
\end{split}
\end{equation}
where the $t_{ij}^{\prime}$ matrix describes the next-nearest neighbor
hopping. The $t^{\prime}$-hopping is an intra-sublattice hopping for each of
the two sublattices ${\bf A}$ and ${\bf B}$ of a bipartite square lattice,
whereas the $t$-hopping is an inter-sublattice hopping.

\begin{figure}[tbp]
\centering
\includegraphics*[width=4.5cm,angle=0]{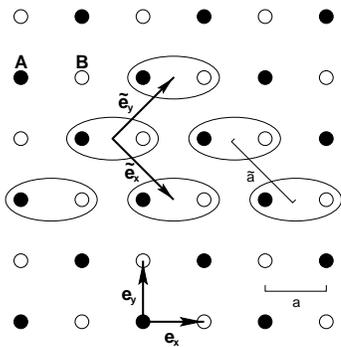}
\caption{The bipartite lattice structure induced by the antiferromagnetic
ordering on a 2 dimensional square lattice.} \label{fig:2}
\end{figure}

In this paper, we consider an antiferromagnetic solution of the
Hamiltonian~(\ref{eq:3.1}) characterized by an opposite sign in the spin
density on nearest neighboring lattice sites. Thus, the antiferromagnetic
ordering induces a magnetic lattice $\{\widetilde{\mathbf R}_{i}\}$ with
lattice constant $\widetilde{a}=a\sqrt{2}$, overlaying the chemical Bravais
lattice $\{{\mathbf R}_{i}\}$ with lattice constant $a$. As shown in
Fig.~\ref{fig:2}, the magnetic lattice is obtained as a square lattice with
basis by collecting together two neighboring sites of the chemical lattice.

During the calculation of the Green's functions it will be convenient to
switch between the two equivalent representations of the system constructed on
the chemical and on the magnetic lattice. We denote vectors belonging to the
chemical lattice by their bare symbols, whereas the corresponding vectors on
the magnetic lattice are denoted by an additional
`$\,\widetilde{\cdot}\,$'-symbol.

We impose the following global boundary conditions on the antiferromagnetic
thermal equilibrium state:
\begin{assumption}
\label{ass:1} The antiferromagnetic thermal equilibrium state has to satisfy
the following global boundary conditions:
\begin{enumerate}
\item The particle and spin densities are globally conserved.
\item The mean value of the local particle density per spin satisfies the relation
\begin{equation}
\label{eq:3.2} \bigl\langle n_\sigma \left( i\right) \bigr\rangle =\frac
12\bigl [n-\left( -1\right) ^\sigma m\cos \left( {\mathbf Q}\cdot {\mathbf
R}_i\right)\bigr ]
\end{equation}
where $m$ is the staggered magnetization, which has to be calculated
self-consistently. The particle density $n$ is imposed as an external
parameter. Further we adopt the convention $\sigma=1,2$ for
$\sigma=\uparrow,\downarrow$ and ${\mathbf Q}=\left (\frac \pi a,\frac \pi
a\right)$.
\item All expectation values are invariant under the contemporary exchange
of the spin direction and the sublattice index.
\item The normalization matrix $I$ and the $m$ matrix from
Eqs.~\eqref{eq:2.6} and \eqref{eq:2.7} have to be real.
\end{enumerate}
\end{assumption}
We remark that the conditions 1.\ -- 3.\ are natural boundary conditions for
an antiferromagnetic thermal equilibrium state characterized by a staggered
magnetization, whereas the condition 4.\ is a direct consequence of the fact
that we are looking for a state without quasi-particle damping, as required by
the approximation scheme described in Sec.~\ref{sec:2}.

To reduce the computational effort in solving the integral self-consistent
equations we use a \emph{spherical approximation} for the
$t^{\prime}$-hopping, characterized by an additional hopping to
next-next-nearest neighbors with half the weight of the next-nearest neighbor
hopping, in the calculations of the integrals in the momentum space. This
permits a definition of the $t^{\prime}$-hopping matrix by $\alpha({\mathbf
k})$ only and therefore reduces the number of evaluations to $O(N)$ -- $N$ the
number of steps in the integrals -- independently from the dimensionality of
the system, while it would have been $O(N^d)$ for a $d$-dimensional system
with the usual next-nearest neighbor $t^{\prime}$-hopping matrix. In $2D$ the
$t^{\prime}$-hopping matrix is given by $t_{ij}^{^{\prime }}=-12t^{\prime}\beta
_{ij}=-12t^{\prime}\frac 1N\sum_{{\mathbf k}}e^{i{\mathbf k\cdot }\left(
{\mathbf R}_i-{\mathbf R}_j\right) }\beta \left( {\mathbf k}\right)$ with
$\beta \left( {\mathbf k}\right) =\frac 13( 4\left[ \alpha \left({\mathbf
k}\right) \right] ^2-1)$. In $3D$ we find $t_{ij}^{\prime}=-30t^{\prime}\beta
_{ij}=-30t^{\prime}\frac 1N\sum_{{\mathbf k}}e^{i{\mathbf k\cdot }\left(
{\mathbf R}_i-{\mathbf R}_j\right) }\beta \left( {\mathbf k}\right)$ with
$\beta \left ( {\mathbf k}\right) =\frac 25( 3\left[ \alpha \left({\mathbf
k}\right) \right] ^2-\frac 12)$. Let us remark that in the present definitions
the hopping per lattice site is always normalized to $1$.

It is worth noting that, as we will take the $t^{\prime}\rightarrow0$ limit
hereafter, the use of the \emph{spherical approximation} can not affect at all
the results we will obtain.

In the following we will restrict our analysis to the $2D$ case. The $3D$
case, which might be treated in complete analogy by renormalizing the hopping
constants and by changing the projections $\alpha_{ij}$ and $\beta_{ij}$, will
be considered in Sec.~\ref{sec:4.3}.

\subsection{The Normalization- and the Energy- Matrices}
\label{sec:3.3}

In order to calculate the energy matrix in the two-pole approximation under
the antiferromagnetic boundary conditions given in Ass.~\ref{ass:1}, we
proceed along the guidelines given in Sec.~\ref{sec:2}. To this end it is
convenient to use the corresponding quantities defined on the chemical lattice
and to switch to the magnetic lattice only for the calculation of the Green's
functions.

Firstly, we calculate the full equation of motion for the composite field
spinor~\eqref{eq:2.2} according to Eq.~\eqref{eq:2.4} for the
$t$--$t^{\prime}$--$U$ Hamiltonian

\end{multicols} \widetext

\begin{equation}
  \label{eq:3.4}
  i\frac \partial {\partial t}\Psi\left( i,t\right) =\left[
    \Psi\left( i,t\right) ,H^{\text{\em tt'U}}\right] = \left(
    \begin{tabular}{l}
      $-\mu \xi _{\uparrow }\left( i,t\right) -4tc_{\uparrow }^\alpha \left(
        i,t\right) -4t\pi _{\uparrow }^\alpha \left( i,t\right) -12t^{^{\prime
          }}c_{\uparrow }^\beta \left( i,t\right) -12t^{\prime}\pi _{\uparrow
        }^\beta \left( i,t\right) $ \\
      $-\mu \eta _{\uparrow }\left( i,t\right) +U\eta _{\uparrow }\left( i,t\right)
      +4t\pi _{\uparrow }^\alpha \left( i,t\right) +12t^{\prime}\pi _{\uparrow
        }^\beta \left( i,t\right) $ \\
      $-\mu \xi _{\downarrow }\left( i,t\right) -4tc_{\downarrow }^\alpha \left(
        i,t\right) -4t\pi _{\downarrow }^\alpha \left( i,t\right) -12t^{^{\prime
          }}c_{\downarrow }^\beta \left( i,t\right) -12t^{\prime}\pi _{\downarrow
        }^\beta \left( i,t\right) $ \\
      $-\mu \eta _{\downarrow }\left( i,t\right) +U\eta _{\downarrow }\left(
        i,t\right) +4t\pi _{\downarrow }^\alpha \left( i,t\right) +12t^{\prime}\pi
      _{\downarrow }^\beta \left( i,t\right) $
    \end{tabular}
  \right)
\end{equation}

\begin{multicols}{2} \narrowtext

\noindent where we used the notation ${\mathbf \pi}_\sigma ^\gamma \left(
i\right) = -n_{\overline{\sigma }}\left( i\right) c_\sigma ^\gamma \left(
i\right) +c_{\overline{\sigma }}^{\dagger }\left( i\right) c_\sigma \left(
i\right) c_{\overline{\sigma }}^\gamma \left( i\right) +c_\sigma \left(
i\right) c_{\overline{\sigma }}^{\gamma \dagger }\left( i\right)
c_{\overline{\sigma }}\left( i\right)$ with $c_\sigma ^\gamma \left( i\right)
=\sum_j\gamma _{ij}c_\sigma \left( j\right)$, where $\gamma =\alpha ,\beta$
stands for the projections defined in Sec.~\ref{sec:3.1}.

According to Ass.~\ref{ass:1}.1, the expectation values for spin exchange
operators vanish and we find the general block structure:

\begin{equation}
  \label{eq:3.5}
  \begin{array}{ccc}
    I=I_{\uparrow }\bigoplus I_{\downarrow } & \quad ,\quad  & \varepsilon
    =\varepsilon _{\uparrow }\bigoplus \varepsilon _{\downarrow } \\[1ex]
    M=M_{\uparrow }\bigoplus M_{\downarrow } & \quad ,\quad  & S=S_{\uparrow
      }\bigoplus S_{\downarrow}\, .
  \end{array}
\end{equation}
The explicit form of the normalization matrix is obtained by direct evaluation
of Eq.~\eqref{eq:2.6}
\begin{equation}
  \label{eq:3.6}
  I\left( i,j\right) =\delta _{ij}I^{\left( n\right) }+\delta _{ij}I^{\left(
      m\right) }\cos \left( {\mathbf Q}\cdot {\mathbf R}_i\right)
\end{equation}
with $I^{\left( n\right) }=\widehat{I}^{\left( n\right) }\bigoplus \widehat{I}
^{\left( n\right) }$ and $I^{\left( m\right) }=\widehat{I}^{\left( m\right)
  }\bigoplus -\widehat{I} ^{\left( m\right) }$ where the blocks are given by

\begin{equation}
  \label{eq:3.7}
  \begin{array}{ccc}
    \widehat{I}^{\left( n\right) }=\left(
      \begin{array}{cc}
        1-\frac n2 & 0 \\
        0 & \frac n2
      \end{array}
    \right)  & \quad ,\quad  & \widehat{I}^{\left( m\right) }=\left(
      \begin{array}{cc}
        \frac m2 & 0 \\
        0 & -\frac m2
      \end{array}
    \right) \, .
  \end{array}
\end{equation}
For the $m$-matrix we get from Def.~\eqref{eq:2.7}
\begin{equation}
  \label{eq:3.8}
  \begin{split}
    m\left( i,j\right) = & \delta _{ij}M_1+\alpha _{ij}M_3+\beta _{ij}M_5\\
    & +\cos\left( \mathbf{Q}\cdot \mathbf{R}_i\right) \left( \delta
    _{ij}M_2+\alpha_{ij}M_4+\beta _{ij}M_6\right) \, .
  \end{split}
\end{equation}
The matrices $M_1,\ldots,M_6$ also have the above mentioned block structure
\begin{equation}
  \label{eq:3.9}
  \begin{array}{lll}
    M_1=\left(
      \begin{array}{ll}
        \widehat{M}_1 & \mathbf{0} \\
        \mathbf{0} & \widehat{M}_1
      \end{array}
    \right)  & \quad ,\quad  & M_2=\left(
      \begin{array}{ll}
        \widehat{M}_2 & \mathbf{0} \\
        \mathbf{0} & -\widehat{M}_2
      \end{array}
    \right)  \\
    M_3=\left(
      \begin{array}{ll}
        \widehat{M}_3 & \mathbf{0} \\
        \mathbf{0} & \widehat{M}_3
      \end{array}
    \right)  & \quad ,\quad  & M_4=\left(
      \begin{array}{ll}
        \widehat{M}_4 & \mathbf{0} \\
        \mathbf{0} & -\widehat{M}_4
      \end{array}
    \right)  \\
    M_5=\left(
      \begin{array}{ll}
        \widehat{M}_5 & \mathbf{0} \\
        \mathbf{0} & \widehat{M}_5
      \end{array}
    \right)  & \quad ,\quad  & M_6=\left(
      \begin{array}{ll}
        \widehat{M}_6 & \mathbf{0} \\
        \mathbf{0} & -\widehat{M}_6
      \end{array}
    \right)
  \end{array} \, .
\end{equation}
with

\end{multicols}

\widetext

\begin{align*}
  \widehat{M}_1 &=\left(
    \begin{array}{cc}
      -\mu \bigl( 1-\frac 12n\bigr) -4t\bigl( \Delta _{\downarrow }^\alpha
      +\Delta _{\uparrow }^\alpha \bigr) -6t^{\prime}\bigl( \Delta
      _{\downarrow }^\beta +\Delta _{\uparrow }^\beta \bigr)  & 4t\bigl( \Delta
      _{\downarrow }^\alpha +\Delta _{\uparrow }^\alpha \bigr) +6t^{^{\prime
          }}\bigl( \Delta _{\downarrow }^\beta +\Delta _{\uparrow }^\beta \bigr) \\[1ex]
      4t\bigl( \Delta _{\downarrow }^\alpha +\Delta _{\uparrow }^\alpha \bigr)
      +6t^{\prime}\bigl( \Delta _{\downarrow }^\beta +\Delta _{\uparrow
        }^\beta \bigr)  & \bigl( U-\mu \bigr) \frac 12n-4t\bigl( \Delta
      _{\downarrow }^\alpha +\Delta _{\uparrow }^\alpha \bigr) -6t^{^{\prime
          }}\bigl( \Delta _{\downarrow }^\beta +\Delta _{\uparrow }^\beta \bigr)
    \end{array}
  \right) \label{eq:3.10}\tag{\theequation}\addtocounter{equation}{1} \\[2ex]
  \widehat{M}_2 &=\left(
    \begin{array}{cc}
      -\mu \frac m2 -4t\bigl( \Delta _{\downarrow }^\alpha -\Delta _{\uparrow }^\alpha
      \bigr) -6t^{\prime}\bigl( \Delta _{\downarrow }^\beta -\Delta
      _{\uparrow }^\beta \bigr)  & 4t\bigl( \Delta _{\downarrow }^\alpha -\Delta
      _{\uparrow }^\alpha \bigr) +6t^{\prime}\bigl( \Delta _{\downarrow
        }^\beta -\Delta _{\uparrow }^\beta \bigr)  \\[1ex]
      4t\bigl( \Delta _{\downarrow }^\alpha -\Delta _{\uparrow }^\alpha \bigr)
      +6t^{\prime}\bigl( \Delta _{\downarrow }^\beta -\Delta _{\uparrow
        }^\beta \bigr)  & -\frac m2 \bigl( -\mu +U\bigr) -4t\bigl( \Delta _{\downarrow
        }^\alpha -\Delta _{\uparrow }^\alpha \bigr) -6t^{\prime}\bigl( \Delta
      _{\downarrow }^\beta -\Delta _{\uparrow }^\beta \bigr)
    \end{array}
  \right) \\[2ex]
  \widehat{M}_3 &=\left(
    \begin{array}{cc}
      -4t\bigl( 1-n+p\bigr)  & -4t\bigl( \frac 12n-p\bigr)  \\[1ex]
      -4t\bigl( \frac 12n-p\bigr)  & -4tp
    \end{array}
  \right) \quad \widehat{M}_5 =\left(
    \begin{array}{cc}
      -12t^{\prime }\bigl( 1-n+\frac 12( p_{\downarrow }^\beta +p_{\uparrow
        }^\beta ) \bigr)  & -12t^{\prime}\bigl( \frac 12n-\frac 12(
      p_{\downarrow }^\beta +p_{\uparrow }^\beta ) \bigr)  \\[1ex]
      -12t^{\prime}\bigl( \frac 12n-\frac 12( p_{\downarrow }^\beta
      +p_{\uparrow }^\beta ) \bigr)  & -6t^{\prime}\bigl( p_{\downarrow
        }^\beta +p_{\uparrow }^\beta \bigr)
    \end{array}
  \right)\\[2ex]
  \widehat{M}_4 &=\left(
    \begin{array}{cc}
      0 & -2tm \\[1ex]
      2tm & 0
    \end{array}
  \right) \quad
  \widehat{M}_6 =\left(
    \begin{array}{cc}
      -12t^{\prime}\bigl( m+\frac 12 ( p_{\downarrow }^\beta -p_{\uparrow
        }^\beta ) \bigr)  & 6t^{\prime}\bigl( m+
      p_{\downarrow }^\beta -p_{\uparrow }^\beta  \bigr)  \\[1ex]
      6t^{\prime}\bigl( m + p_{\downarrow }^\beta -p_{\uparrow
        }^\beta  \bigr)  & -6t^{\prime}\bigl( p_{\downarrow }^\beta
      -p_{\uparrow }^\beta \bigr)
    \end{array}
  \right)\, . \\
\end{align*}
\begin{multicols}{2} \narrowtext To abbreviate the notation of the occurring expectation values we
defined the following parameters
\begin{eqnarray}
\label{eq:3.11} &&\Delta _\sigma ^\alpha =\frac 12\Bigl( \Bigl\langle \xi
_\sigma \left(i\right) c_\sigma ^{\alpha \dagger }\left( i\right) \Bigr\rangle
  -\Bigl\langle c_\sigma ^\alpha \left( i\right) \eta _\sigma ^{\dagger
    }\left( i\right) \Bigr\rangle \Bigr) \,\, ,\,\, \text{for}\,\, i\in
  {\mathbf A}\nonumber\\
   &&p=\frac 14\Bigl\langle n_\mu \left( i\right) n_\mu ^\alpha \left( i\right)
    \Bigr\rangle -\Bigl\langle c_{\uparrow }\left( i\right) c_{\downarrow
      }\left( i\right) \bigl[ c_{\downarrow }^{\dagger }\left( i\right)
    c_{\uparrow }^{\dagger }\left( i\right) \bigr] ^\alpha
    \Bigr\rangle\nonumber\\
&&    \Delta _\sigma ^\beta =\Bigl\langle \xi _\sigma \left( i\right) c_\sigma
    ^{\beta \dagger }\left( i\right) \Bigr\rangle -\Bigl\langle c_\sigma ^\beta
    \left( i\right) \eta _\sigma ^{\dagger }\left( i\right) \Bigr\rangle \,\,
    ,\,\, \text{for}\,\, i\in {\mathbf A}\nonumber\\
&& p_\sigma ^\beta =\frac 14\Bigl\langle \bigl( n_1\left( i\right)
-\left(-\right) ^\sigma in_2\left( i\right) \bigr)\nonumber\\
&&\quad \times \bigl[ n_1\left( i\right) +\left( -\right) ^\sigma i n_2\left(
i\right) \bigr]^\beta \Bigr\rangle + \Bigl\langle n_\sigma \left( i\right)
n_\sigma ^\beta
\left( i\right) \Bigr\rangle\nonumber\\
&&\quad -\Bigl\langle c_{\overline{\sigma }}\left( i\right) c_\sigma \left(
i\right) \bigl[ c_\sigma ^{\dagger }\left( i\right) c_{\overline{\sigma
}}^{\dagger }\left( i\right) \bigr] ^\beta \Bigr\rangle \quad\text{for}\,\,
i\in {\mathbf A}\, .
\end{eqnarray}
Furthermore, we used the notation $n_\mu \left( i\right) = c^{\dagger }\left (
i\right) \sigma _\mu c\left( i\right)$ for the spin- and charge- density
operators with the Pauli spin matrices $\sigma _\mu \in\left\{ {\bf 1},\sigma
_x,\sigma _y,\sigma _z\right\}$ and the electronic field spinors $c^{\dagger
}\left( i\right) = (c^{\dagger}_{\uparrow}(i) \, , \,
c^{\dagger}_{\downarrow}(i))$.

As the antiferromagnetic ordering breaks the translational invariance, the
parameters $\Delta_\sigma^\alpha$, $p$, $\Delta_\sigma^\beta$ and
$p_\sigma^\beta$ in principle do depend on the lattice site $i$. However, the
antiferromagnetic state enjoys a translational invariance within each one of
the two sublattices $\mathbf A$ and $\mathbf B$. For the definitions of the
parameters~\eqref{eq:3.11} we arbitrarily choose the values on the sublattice
$\mathbf A$. Their values on the sublattice $\mathbf B$ are then given by
exchanging the spin indices according to Ass.~\ref{ass:1}.3.

The operators, from which the expectation values for the parameters
$\Delta_\sigma^\alpha$, $p$, $\Delta_\sigma^\beta$ and $p_\sigma^\beta$ are
taken, are not hermitian. However, the corresponding parameters have to be
real according to Ass.~\ref{ass:1}.4. This finally results in the parameter $p$
being independent of both the spin and the sublattice. Furthermore, the
normalization matrix and the $m$-matrix result symmetric.

To calculate the energy matrix we note that Eq.~\eqref{eq:2.5} gives
$m=\varepsilon I$. In the Fourier space we have

\begin{equation}
  \label{eq:3.12}
  \begin{split}
    m\left( {\mathbf k},{\mathbf p}\right)  &= \frac{a^2}{\left( 2\pi \right) ^2}
    \int_{\Omega _B}{\rm d}^2 q\; \varepsilon \left( {\mathbf k+q,p-q}\right) I\left(
      {\mathbf q}\right)  \\
    &= \varepsilon \left( {\mathbf k,p}\right) I^{\left( n\right) }+\varepsilon
    \left( {\mathbf k+Q,p-Q}\right) I^{\left( m\right) }
  \end{split}
\end{equation}
where the Fourier transform of the normalization and the $m$-matrix are given
by
\begin{equation}
  \label{eq:3.13}
  \begin{split}
    I\left( {\mathbf k,p}\right) &= \Bigl( \frac{2\pi
        }a\Bigr) ^2\left( \delta \left( {\mathbf p}\right) I^{\left( n\right)
        }+\delta \left( {\mathbf p}- {\mathbf Q}\right) I^{\left( m\right)
        }\right)\\
    m\left( {\mathbf k},{\mathbf p}\right) &= \Bigl( \frac{2\pi }a\Bigr)
        ^2\left( \vphantom{\sum}\right .\delta \left( {\mathbf p}\right) \left
        ( M_1+\alpha \left( {\mathbf k}\right) M_3+\beta \left( {\mathbf
            k}\right) M_5\right)\\
      & \qquad +\delta \left( {\mathbf p-Q}\right) \left( M_2+\alpha \left
          ( {\mathbf k}\right) M_4+\beta \left( {\mathbf k}\right)
      M_6\right)\left .\vphantom{\sum} \right) \, .
  \end{split}
\end{equation}
We recall that the Fourier transform on the chemical lattice is defined as in
Eq.~\eqref{eq:2.9} to benefit from the periodicity of the thermal equilibrium
states.

Using the $2{\mathbf Q}$ periodicity of the thermal equilibrium states in
Eq.~\eqref{eq:3.12} we get the energy matrix in Fourier space
\begin{equation}
  \label{eq:3.16}
  \varepsilon \left( {\mathbf k,p}\right) =m\left( {\mathbf k},{\mathbf p}\right)
  C-m\left( {\mathbf k+Q},{\mathbf p-Q}\right) D
\end{equation}
with
\begin{equation}
  \label{eq:3.17}
  \begin{array}{l}
    C=\left( I^{\left( m\right) }\right) ^{-1}\bigl[ I^{\left( n\right) }\left(
        I^{\left( m\right) }\right) ^{-1}-I^{\left( m\right) }\left( I^{\left(
            n\right) }\right) ^{-1}\bigr] ^{-1} \\[1ex]
    D=\left( I^{\left( n\right) }\right) ^{-1}\bigl[ I^{\left( n\right) }\left(
        I^{\left( m\right) }\right) ^{-1}-I^{\left( m\right) }\left( I^{\left(
            n\right) }\right) ^{-1}\bigr] ^{-1} \, .
    \end{array}
\end{equation}
Using the explicit expressions~\eqref{eq:3.13} for the $m$- and the
normalization matrix we can write the energy matrix as
\begin{equation}
  \label{eq:3.18}
  \begin{split}
  \varepsilon \left( {\mathbf k,p}\right) &= \Bigl( \frac{2\pi }a\Bigr) ^2\Bigl(
    \delta \left({\mathbf p}\right) \left
    ( \varepsilon ^{(1)}+\alpha \left( {\mathbf k}\right) \varepsilon ^{(2)}+\beta
    \left( {\mathbf k}\right) \varepsilon^{(5)}\right)\\
    & \qquad +\delta \left( {\mathbf p-Q}\right) \left( \varepsilon ^{(3)}+\alpha \left(
        {\mathbf k}\right) \varepsilon ^{(4)}+\beta \left( {\mathbf k}\right) \varepsilon
      ^{(6)}\right)\Bigr)
  \end{split}
\end{equation}
with
\begin{equation}
  \label{eq:3.19}
  \begin{array}{ccc}
    \varepsilon ^{(1)}= M_1C-M_2D  & \quad ,\quad  & \varepsilon
    ^{(2)}= M_3C+M_4D  \\
    \varepsilon ^{(3)}= M_2C-M_1D  & \quad ,\quad  & \varepsilon
    ^{(4)}= M_4C+M_3D  \\
    \varepsilon ^{(5)}= M_5C-M_6D  & \quad ,\quad  & \varepsilon
    ^{(6)}= M_6C-M_5D \, .
  \end{array}
\end{equation}

\subsection{The Green's Functions}
\label{sec:3.4}

The solutions of the linearized equation of motion~\eqref{eq:2.10} with the
expression~\eqref{eq:3.18} for the energy matrix can be interpreted as
translational invariant Green's functions $S^{AA}(\widetilde{{\mathbf
    k}},\omega)$, $S^{AB}(\widetilde{{\mathbf k}},\omega)$,
$S^{BA}(\widetilde{{\mathbf k}},\omega)$ and $S^{BB}(\widetilde{{\mathbf
    k}},\omega)$ on the magnetic lattice as is illustrated in
App.~\ref{sec:A.1}. They have the general structure
\begin{equation}
  \label{eq:3.21}
  \begin{split}
    S^{XY}( \widetilde{\mathbf{k}},\omega)&=\left( \omega
      ^2+\omega A^{XY}( \widetilde{{\mathbf k}}) +B^{XY}
      ( \widetilde{{\mathbf k}}) \right)^{-1}\\
    &\quad\times\left( \omega C^{XY}(\widetilde{{\mathbf k}})
      +D^{XY}(\widetilde{{\mathbf k }}) \right)
  \end{split} \, .
\end{equation}
The explicit form of the coefficients $A^{XY}$, $B^{XY}$, $C^{XY}$ and
$D^{XY}$ are given in App.~\ref{sec:A.2}.

Due to the assumptions~\ref{ass:1} on the antiferromagnetic thermal
equilibrium state, the Green's functions~\eqref{eq:3.21} have the block
structure $S^{XY}=S^{XY}_{\uparrow}\bigoplus S^{XY}_{\downarrow}$ shown in
Eq.~\eqref{eq:3.5}, where the poles $\{E_{\sigma,i}^{XY}( \widetilde{{\mathbf
    k}})\, \vert \, i\in\{1,2,3,4\}\}$ of the spin dependent parts
$S^{XY}_{\sigma}$ of the Green's functions are given by the roots of the
fourth order equations
\begin{equation}
  \label{eq:3.25}
  \det \left(\omega ^2+\omega A_\sigma ^{XY}( \widetilde{{\mathbf k}})
    +B_\sigma ^{XY}( \widetilde{{\mathbf k}}) \right) =0 \, .
\end{equation}
The set of poles for the Green's functions $S^{XY}_{\sigma}$ are all equal,
i.e.\ the quasi-particle energies do depend neither on the spin nor on the
sublattice. This reflects the property of the antiferromagnetic state with
staggered magnetization, where the majority spin states and the minority spin
states energetically occupy exactly the same regions and differ only in their
corresponding spectral weights (cf.\ also Sec.~\ref{sec:4}). Therefore, we
simply write for the poles $E_i(\widetilde{{\mathbf k}}) = E_{\sigma,i}
^{XY}(\widetilde{{\mathbf k}})$.

We can write the retarded Green's functions~\eqref{eq:3.21} as
\begin{equation}
  \label{eq:3.26}
  S_\sigma ^{XY}( \widetilde{{\mathbf k}},\omega )
  =\lim\limits_{\eta \rightarrow 0}\sum_{i=1}^4\frac 1{\omega
  -E_{i}( \widetilde{{\mathbf k}}) +i\eta }\sigma _{\sigma
    ,i}^{XY}( \widetilde{{\mathbf k}})
\end{equation}
with $X,Y\in \left\{{\mathbf A},{\mathbf B}\right\}$, $\sigma \in \left\{
\uparrow
  ,\downarrow \right\}$ and the spin- and sublattice-dependent spectral
weights $\sigma _{\sigma ,i}^{XY}$ given by \end{multicols} \widetext
\begin{equation}
  \label{eq:3.27}
  \begin{split}
    \sigma _{\sigma ,i}^{XY}( \widetilde{{\mathbf k}})  = \frac
    1{\prod\limits_{
        \begin{subarray}{l}
          j=1\\  j\neq i
        \end{subarray}
        }^4\bigl( E_{i}( \widetilde{{\mathbf k}}) -E_{j}( \widetilde{{\mathbf k}
        }) \bigr) }&
    \left[ \bigl( E_{i}( \widetilde{{\mathbf k}}
      ) \bigr) ^3C_\sigma ^{XY}( \widetilde{{\mathbf
          k}})
      +\bigl( E_{i}( \widetilde{{\mathbf k}}) \bigr) ^2\Bigl(
      D_\sigma ^{XY}( \widetilde{{\mathbf k}}) +\det \bigl( A_\sigma
      ^{XY}( \widetilde{{\mathbf k}}) \bigr) \bigl( A_\sigma
      ^{XY}( \widetilde{{\mathbf k}}) \bigr) ^{-1}C_\sigma ^{XY}(
      \widetilde{{\mathbf k}}) \Bigr)\right .\\[-5ex]
    &\quad +E_{i}( \widetilde{
      {\mathbf k}}) \Bigl( \det \bigl( A_\sigma ^{XY}( \widetilde{
      {\mathbf k}}) \bigr) \bigl( A_\sigma ^{XY}
    ( \widetilde{{\mathbf k}} ) \bigr) ^{-1}D_\sigma
    ^{XY}( \widetilde{{\mathbf k}}) \Bigr)\\
    &\quad \left . +\det \bigl( B_\sigma ^{XY}( \widetilde{{\mathbf
          k}}) \bigr)
      \bigl( B_\sigma ^{XY}( \widetilde{{\mathbf k}}) \bigr)
      ^{-1}C_\sigma ^{XY}( \widetilde{{\mathbf k}}) +\det \bigl(
      B_\sigma ^{XY}( \widetilde{{\mathbf k}}) \bigr) \bigl( B_\sigma
      ^{XY}( \widetilde{{\mathbf k}}) \bigr) ^{-1}D_\sigma ^{XY}(
      \widetilde{{\mathbf k}}) \right] \, .
  \end{split}
\end{equation}
\begin{multicols}{2} \narrowtext

\subsection{The Self Consistency Equations}
\label{sec:3.5}

Given the temperature $T$ and the particle density $n$ as external
thermodynamic parameters as well as the Coulomb interaction $U$ and the
hopping constants $t$ and $t^{\prime}$, which are the model dependent
parameters, we are now able to give a closed set of self-consistent conditions
for the internal parameters characterizing an antiferromagnetic thermal
equilibrium state. These internal parameters are $p$, $p_{\uparrow }^\beta$,
$p_{\downarrow }^\beta$, $\Delta _{\uparrow }^\alpha$, $\Delta _{\downarrow
}^\alpha$, $\Delta _{\uparrow }^\beta$ and $\Delta _{\downarrow }^\beta$ from
Eq.~\eqref{eq:3.11}, the chemical potential $\mu$ and the magnetization $m$
from Eq.~\eqref{eq:3.2}.

For the calculations of these parameters we need the knowledge of the
correlation functions, which are connected to the retarded Green's functions
of the fundamental spinor by means of the spectral theorem. In view of the
special form~\eqref{eq:3.26} of the retarded Green's functions the spectral
theorem at equal time may be written as
\begin{equation}
  \label{eq:3.28}
  \begin{split}
    &C^{XY}( \widetilde{{\mathbf R}}_i,\widetilde{{\mathbf R}}_j)
    =\left\langle {\mathbf \Psi }^X( \widetilde{{\mathbf R}}_i)
      {\mathbf \Psi }^{Y\dagger }( \widetilde{{\mathbf R}}_j)
    \right\rangle \\
    =&\frac 12\frac{\widetilde{a}^2}{\left( 2\pi
      \right) ^2}\sum_{l=1}^4\int_{\Omega _{\widetilde{B}}}{\rm
    d}^2\widetilde{k} \, e^{i\widetilde{
        {\mathbf k}}{\mathbf \cdot }( \widetilde{{\mathbf R}}_i-\widetilde{{\mathbf
            R}}_j) }\\
    &\qquad\qquad\qquad\qquad\times\sigma _l^{XY}( \widetilde{{\mathbf k}})
    \Bigl[ 1+\tanh \Bigl( \frac{E_l ( \widetilde{{\mathbf k}}) }{2k_BT}\Bigr)
    \Bigr] \, .
  \end{split}
\end{equation}
We denote the on-site, the nearest neighbor and the next-(next)-nearest
neighbor correlation functions at equal time by $C^{XX}( \widetilde{{\mathbf
R}}_i)$, $C^{XY\widetilde{\alpha }} ( \widetilde{{\mathbf R}}_i)$ and
$C^{XX\widetilde{\beta }} ( \widetilde{{\mathbf R}}_i)$.

For the parameters $\mu$ and $m$ we then find the self-consistent equations
\begin{subequations}
  \label{eq:3.30}
  \begin{equation}
    \begin{split}
      \label{eq:3.30a}
      2-n&=C_{11}^{AA}( \widetilde{{\mathbf R}}_i)
      +C_{11}^{BB}( \widetilde{{\mathbf R}}_i) +C_{22}^{AA}(
      \widetilde{{\mathbf R}}_i) +C_{22}^{BB}( \widetilde{{\mathbf R}}_i)\\
      2m&=C_{44}^{AA}( \widetilde{{\mathbf R}}_i)
      -C_{22}^{AA}( \widetilde{{\mathbf R}}_i) +C_{22}^{BB}(
      \widetilde{{\mathbf R}}_i) -C_{44}^{BB}( \widetilde{{\mathbf R}}_i) ] \, .
    \end{split}
  \end{equation}
  The parameters $\Delta _{\uparrow }^\alpha$, $\Delta _{\downarrow }^\alpha$,
  $\Delta _{\uparrow }^\beta$ and $\Delta_{\downarrow }^\beta$ are directly
  related to matrix elements of the correlation functions by
  \begin{equation}
    \label{eq:3.30b}
    \begin{split}
      \Delta _{\uparrow }^\alpha &=\frac 12\bigl[ C_{11}^{AB\widetilde{
          \alpha }}( \widetilde{{\mathbf R}}_i) +C_{12}^{AB\widetilde{\alpha
          }}( \widetilde{{\mathbf R}}_i)\\
      &\qquad\quad-C_{12}^{BA\widetilde{\alpha }
        }( \widetilde{{\mathbf R}}_i) -C_{22}^{BA\widetilde{\alpha }
        }( \widetilde{{\mathbf R}}_i) \bigr]  \\
      \Delta _{\downarrow }^\alpha &=\frac 12\bigl[ C_{33}^{AB\widetilde{
          \alpha }}( \widetilde{{\mathbf R}}_i) +C_{34}^{AB\widetilde{\alpha
          }}( \widetilde{{\mathbf R}}_i)\\
      &\qquad\quad -C_{34}^{BA\widetilde{\alpha }
        }( \widetilde{{\mathbf R}}_i)-C_{44}^{BA\widetilde{\alpha }
        }( \widetilde{{\mathbf R}}_i) \bigr]  \\
      \Delta _{\uparrow }^\beta &=\bigl[ C_{11}^{AA\widetilde{\beta }
        }( \widetilde{{\mathbf R}}_i) -C_{22}^{AA\widetilde{\beta }}(
      \widetilde{{\mathbf R}}_i) \bigr]  \\
      \Delta _{\downarrow }^\beta &=\bigl[ C_{11}^{BB\widetilde{\beta
          }}( \widetilde{{\mathbf R}}_i) -C_{22}^{BB\widetilde{\beta }
        }( \widetilde{{\mathbf R}}_i) \bigr] \, .
    \end{split}
  \end{equation}
The parameters $p$, $p_{\uparrow }^\beta$ and $p_{\downarrow }^\beta$ cannot
be calculated explicitly by the single-particle Green's
function~\eqref{eq:3.26}, because they derive from higher-order correlation
functions. According to what stated in Sec.~\ref{sec:2}, we will use the
following equations to fix the representation of the Green's function
\cite{Mancini:95,Mancini:98}
\begin{equation}
  \label{eq:3.30c}
  \begin{split}
    C_{12}^{AA}( \widetilde{{\mathbf R}}_i) &=0 \\
    C_{12}^{BB}( \widetilde{{\mathbf R}}_i) &=0 \\
    C_{11}^{AA}( \widetilde{{\mathbf R}}_i) &=C_{33}^{AA}(
      \widetilde{{\mathbf R}}_i) \, .
  \end{split}
\end{equation}
\end{subequations}
We observe that all the self-consistent equations are coupled and have to be
solved as one set by means of a global convergency scheme.

Finally we remark that the correlation functions $C^{XX}( \widetilde{{\mathbf
    R}}_i)$, $C^{XY\widetilde{\alpha }}( \widetilde{{\mathbf R}}_i)$ and
$C^{XX\widetilde{\beta }}( \widetilde{{\mathbf R}}_i)$ actually do not depend
on the lattice site ${\mathbf R}_i$ of the magnetic lattice because of the
translational invariance enjoined by the Green's functions~\eqref{eq:3.26}.

\section{Numerical Evaluation of the Antiferromagnetic Phase for the Hubbard Model}
\label{sec:4}

In the following the thermodynamics of the antiferromagnetic thermal
equilibrium states for the Hubbard model in two and three dimensions are
discussed. We report the phase diagrams resulting from solutions of
Eqs.~\eqref{eq:3.30} as well as some of the microscopic properties of the
corresponding solutions, which explain the phase behavior. In this context,
our main concern is in the interplay of the antiferromagnetic
(Mott-Heisenberg) gap and the Mott-Hubbard gap leading to a metal-insulator
transition in the antiferromagnetic phase. Furthermore, we investigate the
distribution of spectral weight between the majority and minority spin states
that explains most of the phase properties found for the antiferromagnetic
solution.

As a first step, we establish an averaging procedure between solutions for
positive and negative values of $t^{\prime}$ that is capable to rule out a non
physical artifact induced by the spherical approximation and to establish the
correspondence to the simple Hubbard model.

\subsection{The Averaging Procedure in \lowercase{$\mathbf t^{\prime}$}}
\label{sec:4.1}

\begin{figure}[tbp]
  \centering
  \subfigure[Sublattice Magnetization as Function of $n$]
  {\label{fig:3a}
    \includegraphics*[width=5cm,angle=270]{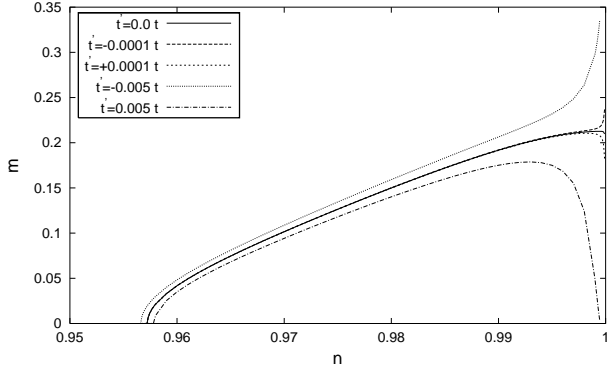}}
  \qquad
  \subfigure[Chemical Potential as Function of $n$]
  {\label{fig:3b}
    \includegraphics*[width=5cm,angle=270]{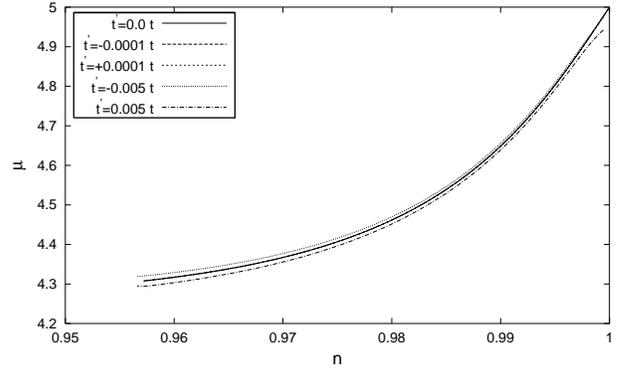}}\\
  \subfigure[The Parameter $p$ as Function of $n$]
  {\label{fig:3c}
    \includegraphics*[width=5cm,angle=270]{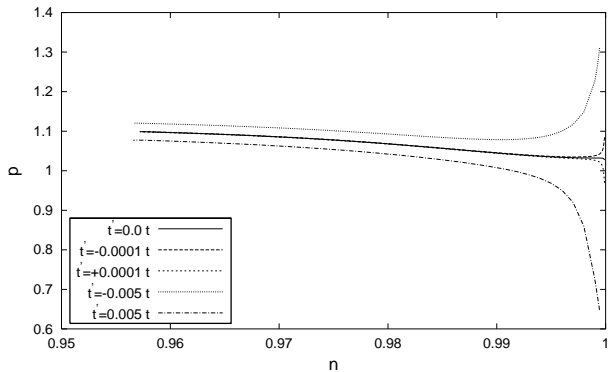}}
  \qquad
  \subfigure[The Parameters $\Delta^\alpha$ as Function of $n$]
  {\label{fig:3d}
    \includegraphics*[width=5cm,angle=270]{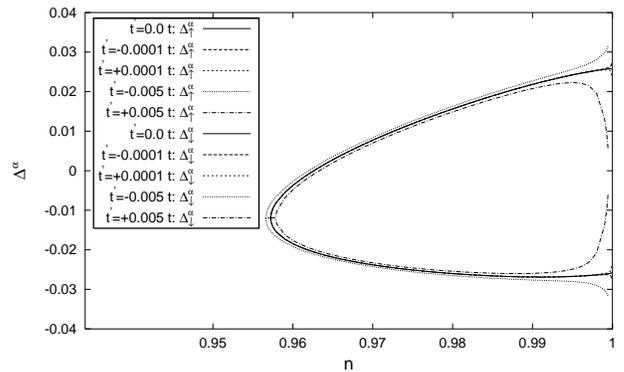}}
  \caption{The averaging algorithm between positive and negative values of
    $t^{\prime}$ in the $2D$ $t$--$t^{\prime}$--$U$ model for $U=10t$ and
    $kT=0.5t$}
  \label{fig:3}
\end{figure}

Due to the incompatibility of the nearest and the next-nearest neighbor
hopping the $t$--$t^{\prime}$--$U$ model does not enjoy the particle-hole
symmetry at half filling ($n=1$). In addition, the spherical approximation for
the $t^{\prime}$-hopping overemphasizes the intra-sublattice hopping to the
next- and next-next-nearest neighbors. This leads to an instability of the
antiferromagnetic solution for the $t$--$t^{\prime}$--$U$ model at half
filling: the magnetization goes to zero for positive values of the
$t^{\prime}$-hopping while for negative values of $t^{\prime}$ diverges to
infinity (cf.\ Fig.~\ref{fig:3a}). Positive values of the intra-sublattice
hopping do suppress antiferromagnetism to a certain extent, whereas negative
values are in favor of it by the additional phase factor of
$\pi$. \\
For the simple Hubbard model the particle-hole symmetry reflects in the
algebraic property
\begin{subequations}
  \label{eq:4.1}
  \begin{equation}
    \label{eq:4.1a}
    \mu(n=1)=\frac U2 \, .
  \end{equation}
  For an antiferromagnetic solution in the framework of the two-pole
  approximation we have the additional
  condition
  \begin{equation}
    \label{eq:4.1b}
    \Delta^{\alpha}_{\uparrow}(n=1) = -\Delta^{\alpha}_{\downarrow}(n=1)\, ,
  \end{equation}
  which is a generalization of the condition $\Delta^{\alpha}_{\uparrow}(n=1) = 0 =
  \Delta^{\alpha}_{\downarrow}(n=1)$, imposed by the particle-hole symmetry on
  the paramagnetic solution of the simple Hubbard model, because of the
  inequivalence of the majority and the minority spin subsystems in the
  antiferromagnetic state.
\end{subequations}
The condition~\eqref{eq:4.1a} is satisfied in the limit $t^{\prime}$ going to
zero independently of the direction (see Fig.~\ref{fig:3b}). This is a direct
consequence of the fact that the proper representation for the Green's
functions has been taken by using the complete set\cite{Avella:98,Mancini:00}
of constrains coming form the Pauli principle~\eqref{eq:3.30c}.

Taking the average between the solutions for $t^{\prime}$ and $-t^{\prime}$
with $\left \vert t^{\prime}\right \vert$ approaching zero we get an
antiferromagnetic state -- which formally has zero $t^{\prime}$-hopping --
that satisfies Eqs.~\eqref{eq:4.1} and has no divergence in the magnetization
and the other parameters at half filling (see Figs.~\ref{fig:3a}, \ref{fig:3c}
and \ref{fig:3d}) and can be considered as representative for the simple
Hubbard model. Averaging the solutions for $\pm t^{\prime}$ thus combines the
large benefit in computational time provided by the spherical approximation
with an antiferromagnetic solution that satisfies the complete set of symmetry
constrains deriving from the Pauli principle and enjoys the particle-hole
symmetry. The numerically accessible limit for the $t^{\prime}$-hopping, which
has been used in combination with the above averaging procedure (see
Fig.~\ref{fig:3}), is $t^{\prime}=\pm 10^{-4}t$. Hereafter, we will present
results exclusively from the averaged solution and we will consider them as
those of the simple Hubbard model.

\end{multicols} \widetext
\begin{multicols}{2} \narrowtext

\subsection{The Antiferromagnetic State of the Hubbard Model}
\label{sec:4.2}

\subsubsection{The Phase Diagram}
\label{sec:4.2.1}

\begin{figure}[tbp]
  \centering
  \subfigure[$U=10t$]
  {\label{fig:5a}
    \includegraphics*[width=5.2cm,angle=270]{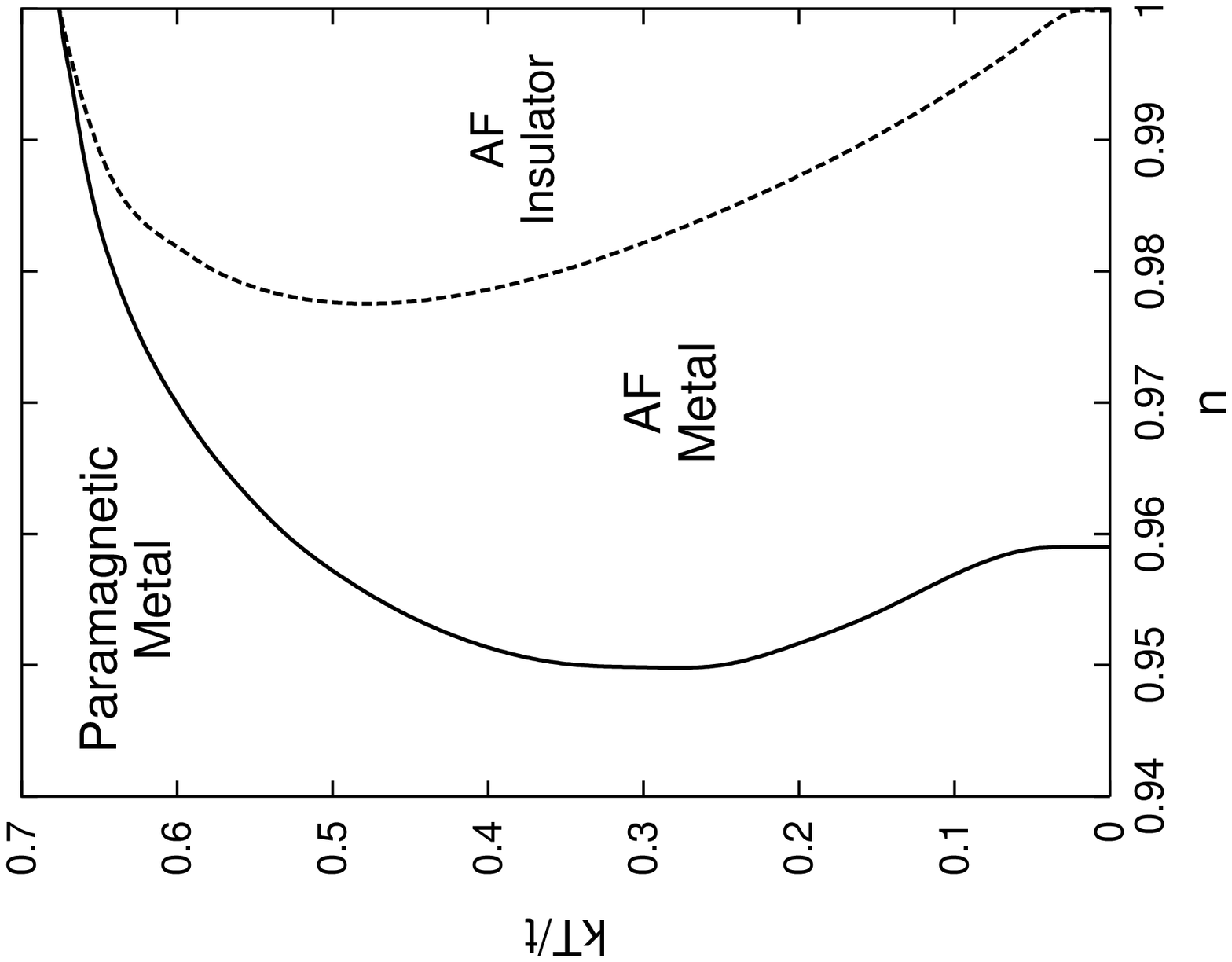}}
  \hskip 1mm
  \subfigure[$U=20t$]
  {\label{fig:5c}
    \includegraphics*[width=5.2cm,angle=270]{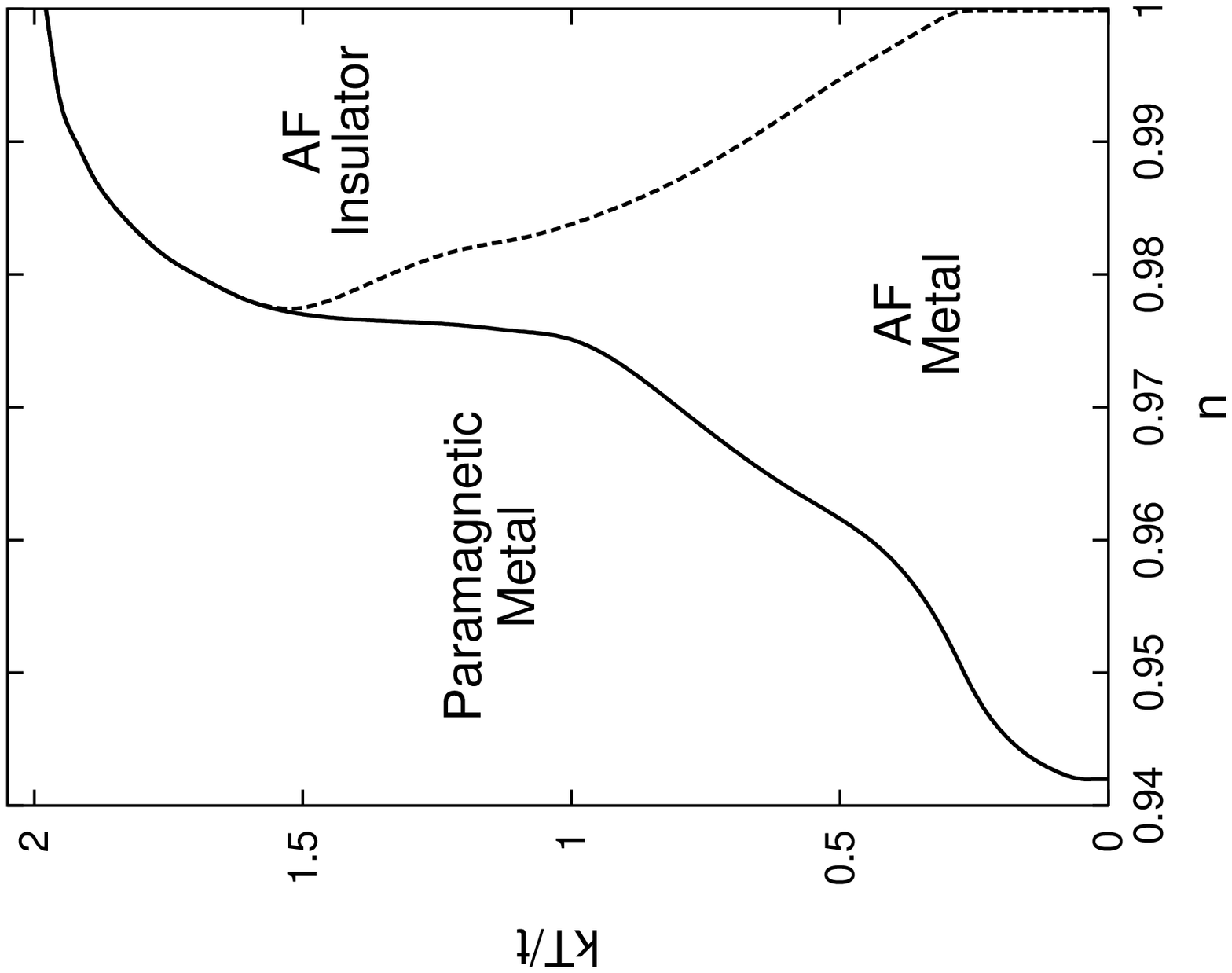}}
  \caption{The $n$--$T$ Phase diagram for the 2D Hubbard model.}
  \label{fig:5}
\end{figure}

\begin{figure}[tbp]
  \centering
  \subfigure[$kT=0.01t$]
  {\label{fig:6a}
    \includegraphics*[width=5.2cm,angle=270]{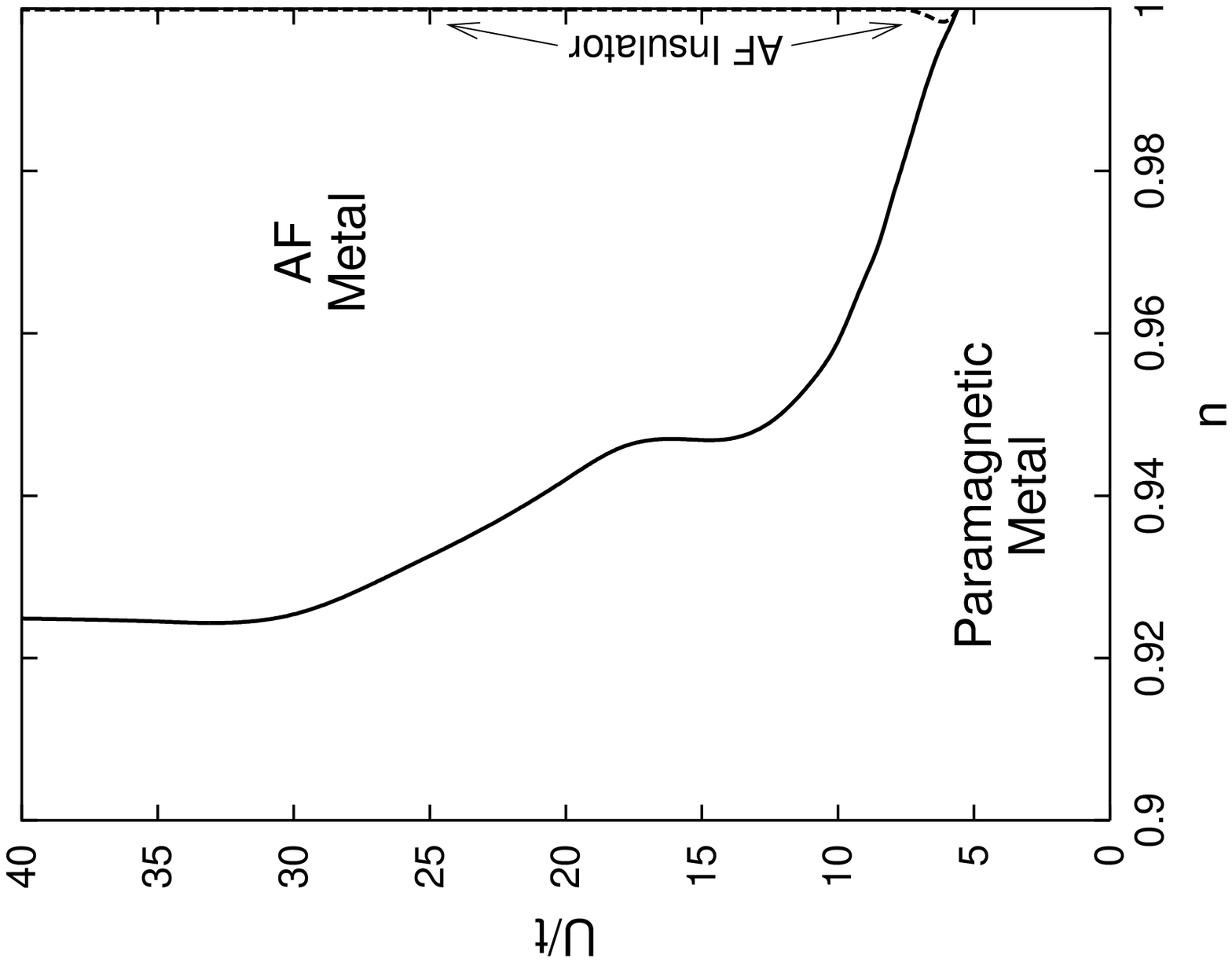}}
  \hskip 1mm
  \subfigure[$kT=0.5$]
  {\label{fig:6c}
    \includegraphics*[width=5.2cm,angle=270]{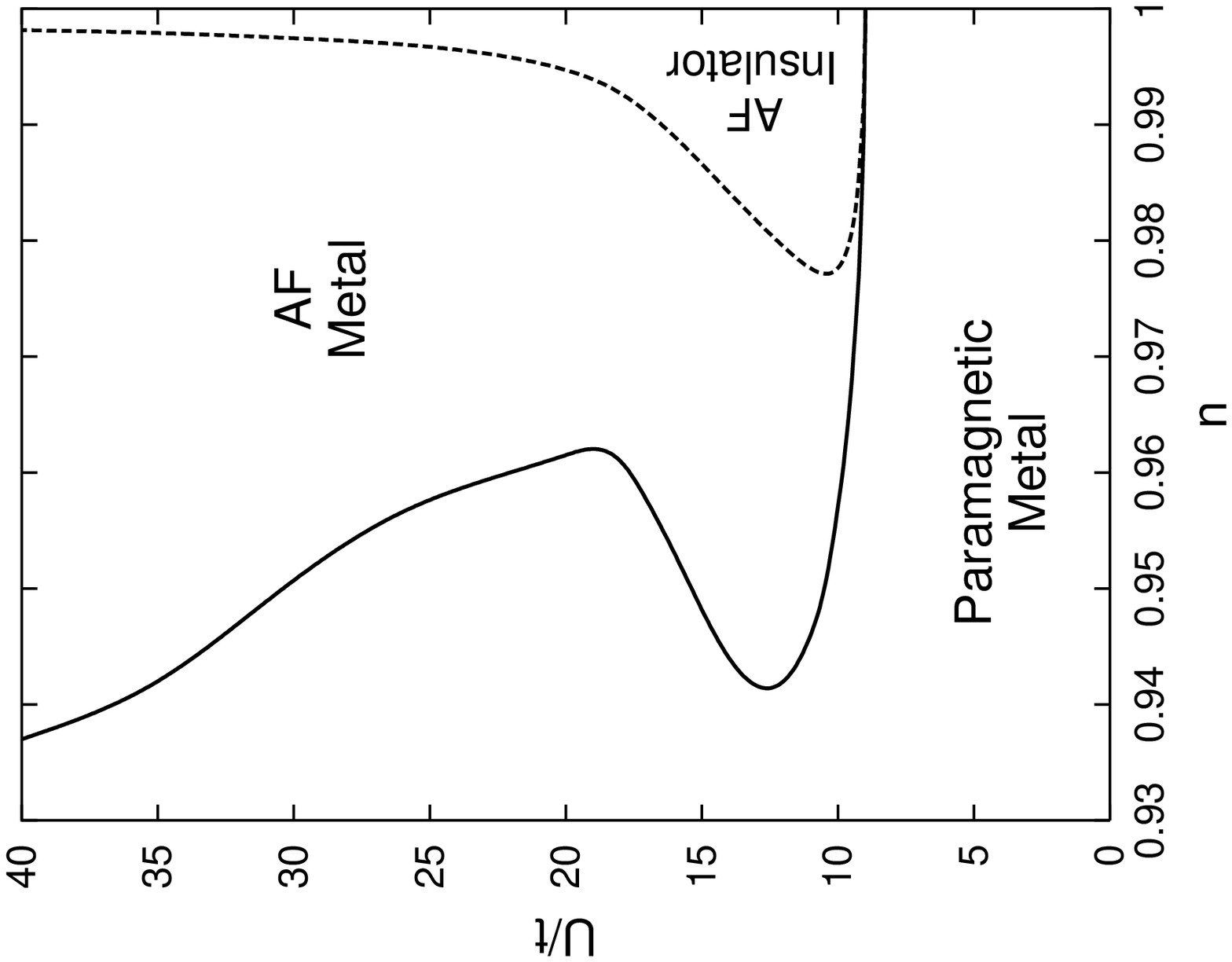}}
  \caption{The $U$--$n$ Phase diagram for the 2D Hubbard model.}
  \label{fig:6}
\end{figure}

The $n$--$T$ and $U$--$n$ phase diagrams for the antiferromagnetic state of
the $2D$ Hubbard model and the corresponding paramagnetic state are shown in
Figs.~\ref{fig:5} and \ref{fig:6}. The antiferromagnetic state has a free
energy lower than the one of the paramagnetic state over the whole phase
region, leading to a phase transition of second order between the
antiferromagnetic and the paramagnetic phase at the lines of vanishing
magnetization. The study of the phase diagram near half filling has to be
completed by an investigation of the ferromagnetic phase \cite{Avella:00a} and
the charge ordered phase, which both can be studied in the framework of the
approximation scheme described above. The antiferromagnetic phase could be
energetically ruled out by one of these other phases in certain regions of the
phase diagrams, then leading to phase transitions of first order
\cite{Herrmann:97a,Chattopadhyay:97,Zhang:92,Zhang:92a}.

In this section we give a brief overview of the properties of the
antiferromagnetic phase, which then will be related to the inner structure of
the antiferromagnetic state -- namely its density of states -- in the
subsequent sections.

The most striking features of the antiferromagnetic phase are the finite
critical Coulomb interaction $U_c$ as a lower bound to the antiferromagnetic
state at half filling, the restriction of the antiferromagnetic state to a
very narrow region in $n$ around half filling and the metal-insulator
transition (MIT) within the antiferromagnetic phase. The vanishing of the
staggered magnetization at half filling for $U<U_c$ -- we find $U_c$ within
$5t$--$10t$ (see Fig.~\ref{fig:6}) -- is supposed to be an effect of strong
electron correlations. It cannot be observed within simple mean-field
treatments of the Hubbard model \cite{Hirsch:85,Oles:81,Trapper:97}, where the
antiferromagnetic phase is stable down to $U=0$ at $n=1$. Also in some more
sophisticated mean-field approximations, as the \emph{SDA}, a stable
antiferromagnetic state is found at $n=1$ down to very small values of $U$ and
its stability down to $U=0$ cannot be excluded \cite{Herrmann:97a}. The same
holds for the antiferromagnetic state of the simple Hubbard model treated by
the \emph{DMFT} \cite{Jarrell:93}.

As we already mentioned in Sec.~\ref{sec:1}, the \emph{SDA} is rather closely
related to the \emph{COM}, but to calculate the antiferromagnetic state
further approximations on the correlation functions were needed
\cite{Kellen:90,Herrmann:97a}. The main difference lies in the treatment of
the internal parameters emerging from higher-order correlation functions.
While the \emph{COM} uses them to fix the representation of the Green's
functions~\eqref{eq:3.30c}, they are calculated by the equations of motion in
the \emph{SDA}.

In a previous study \cite{Munzner:00} we investigated the critical Coulomb
interaction as a function of the $t^{\prime}$-hopping, showing that a finite
value of $U_c$ in the above mentioned range has also to be expected in the
exact limit $t^{\prime}\rightarrow 0$.


The antiferromagnetic phase found in the high-$T_c$ copper oxide compounds in
general shows a great stability at half filling, whereas already a few percent
of electron or hole doping leads to a strong reduction of the N{\'e}el
temperature and eventually to the vanishing of the antiferromagnetic phase. On
the other hand it is well known that in general the mean-field treatment of the
Hubbard model strongly overemphasizes the stability of the antiferromagnetic
phase in doping. In the \emph{COM}, where the strong correlation effects are
restored to some extent by the Pauli principle symmetry constraint, the
stability of the antiferromagnetic phase is actually reduced to a narrow
region of a few percent of doping around half filling with a strong reduction
of the N{\'e}el temperature. For the Hubbard model such a behavior has been
confirmed by numerical results in quantum Monte Carlo studies
\cite{Dagotto:92a,Dagotto:94}. In contrast a larger region for the
antiferromagnetic phase has been obtained by the \emph{SDA}
\cite{Herrmann:97a} and the $d=\infty$ approximation
\cite{Jarrell:93,Georges:96}.

At zero temperature a transition from an antiferromagnetic insulator at half
filling and an antiferromagnetic metal for $n<1$ is observed. At higher
temperatures, however, we find an extended region in doping around $n=1$,
where the Fermi level is situated inside the Mott-Heisenberg gap, which itself
is large with respect to the thermal energy $kT$. We thus have a state with
poor conductivity of semiconductor type, which we call the `antiferromagnetic
insulator'. The metallic behavior is recovered when the Fermi level joins the
second antiferromagnetic band with decreasing values of $n$.

The $n$--$T$ phase diagrams show a strong qualitative difference in the low
temperature region according to the presence (Fig.~\ref{fig:5c}) or the
absence (Figs.~\ref{fig:5a}) of a Mott-Hubbard gap. For values of $U$ that do
not admit a Mott-Hubbard gap the stability of the antiferromagnetic phase is
enhanced by increasing temperature (a phenomenon called `heat magnetization'),
whereas the contrary is true for values of $U$ where the Mott-Hubbard gap is
already opened.

Furthermore, the presence of the Mott-Hubbard gap and the fact that it closes
within the antiferromagnetic phase, when the particle density is reduced, is
also responsible for the intermediate reduction of the stability of the
antiferromagnetic state by increasing the Coulomb interaction
(Fig.~\ref{fig:6}). This reduction is reinforced by increasing temperature and
leads to a `nose-like' shape of the $U$--$n$ phase diagram. Both phenomena can
be explained by the evolution of the spectral weights for the majority and
minority spin subsystems as functions of the external parameters, as will be
explained in detail in the following section.

Finally, we remark that the phase diagrams shown in Figs.~\ref{fig:5} and
\ref{fig:6} are completely symmetric with respect to $n=1$; this is due to the
fact that the thermal equilibrium state respects the particle-hole symmetry. To
incorporate the experimentally observed asymmetry in particle- and hole-doping
a projection of the two-band Hubbard model on an effective single-band one has
been proposed \cite{Zhang:92a}.

\subsubsection{The Band Properties}
\label{sec:4.2.2}

The spectral properties of the antiferromagnetic state are deduced from the
electronic single-particle Green's function on each sublattice
$S^{XX}_{cc\dagger } ( \widetilde{\mathbf k},\omega )$ with $X\in\{{\mathbf
  A}, {\mathbf B}\}$. In the following we will restrict our analysis to the sublattice
$\mathbf A$, while the quantities on the sublattice $\mathbf B$ are obtained
by simply exchanging the majority and minority spin subsystems. We thus have
for the retarded electronic Green's function on the sublattice $\mathbf A$
\begin{equation}
  \label{eq:4.2}
  \begin{split}
    S^{AA}_{cc\dagger }( \widetilde{\mathbf k},\omega )
    =&S^{AA}_{\xi \xi \dagger }( \widetilde{\mathbf k},\omega )
    +S^{AA}_{\xi \eta \dagger }( \widetilde{\mathbf k},\omega
    )\\
    &+S^{AA}_{\eta \xi \dagger }( \widetilde{\mathbf k},\omega
    ) +S^{AA}_{\eta \eta \dagger }( \widetilde{\mathbf
      k},\omega ) \, .
  \end{split}
\end{equation}
The spin dependent electronic density of states within the antiferromagnetic
state is then given by
\begin{equation}
  \label{eq:4.3}
  \begin{split}
    N^A_\sigma \left( \omega \right) :&=\Bigl( \frac{\widetilde{a}}{2\pi
      }\Bigr)^2\int_{\Omega_{\widetilde{B}}}\text{d}^2\widetilde{k}\;\Bigl
    ( -\frac 1\pi \text{Im}S^{AA}_{cc\dagger} ( \widetilde{\mathbf k},\omega)
    \Bigr)\\
    &=\Bigl( \frac{\widetilde{a}}{2\pi }\Bigr)
    ^2\int_{\Omega_{\widetilde{B}}}\text{d}^2\widetilde{k}\;\sum_{i=1}^{4}\delta
    ( \omega -E_{i}( \widetilde{\mathbf k})) \\
    &\qquad\qquad \times \Bigl( \sigma _{i, \xi _\sigma \xi _\sigma ^{\dagger
        }}^{AA}(\widetilde{\mathbf k},\omega) +\sigma _{i,\xi _\sigma \eta _\sigma
      ^{\dagger }}^{AA}( \widetilde{\mathbf k},\omega)\\
    &\qquad\qquad\quad +\sigma _{i,\eta _\sigma \xi _\sigma
      ^{\dagger }}^{AA}( \widetilde{\mathbf k},\omega) +\sigma _{i,\eta _\sigma \eta
      _\sigma ^{\dagger }}^{AA}( \widetilde{\mathbf k},\omega ) \Bigr) \, .
  \end{split}
\end{equation}
We recall that all energies are referred to the chemical potential.

\begin{figure}[tbp]
  \centering
  \subfigure[Sublattice Density of States for $U=16t$]
  {\label{fig:8a}
    \includegraphics*[width=6.4cm,angle=270]{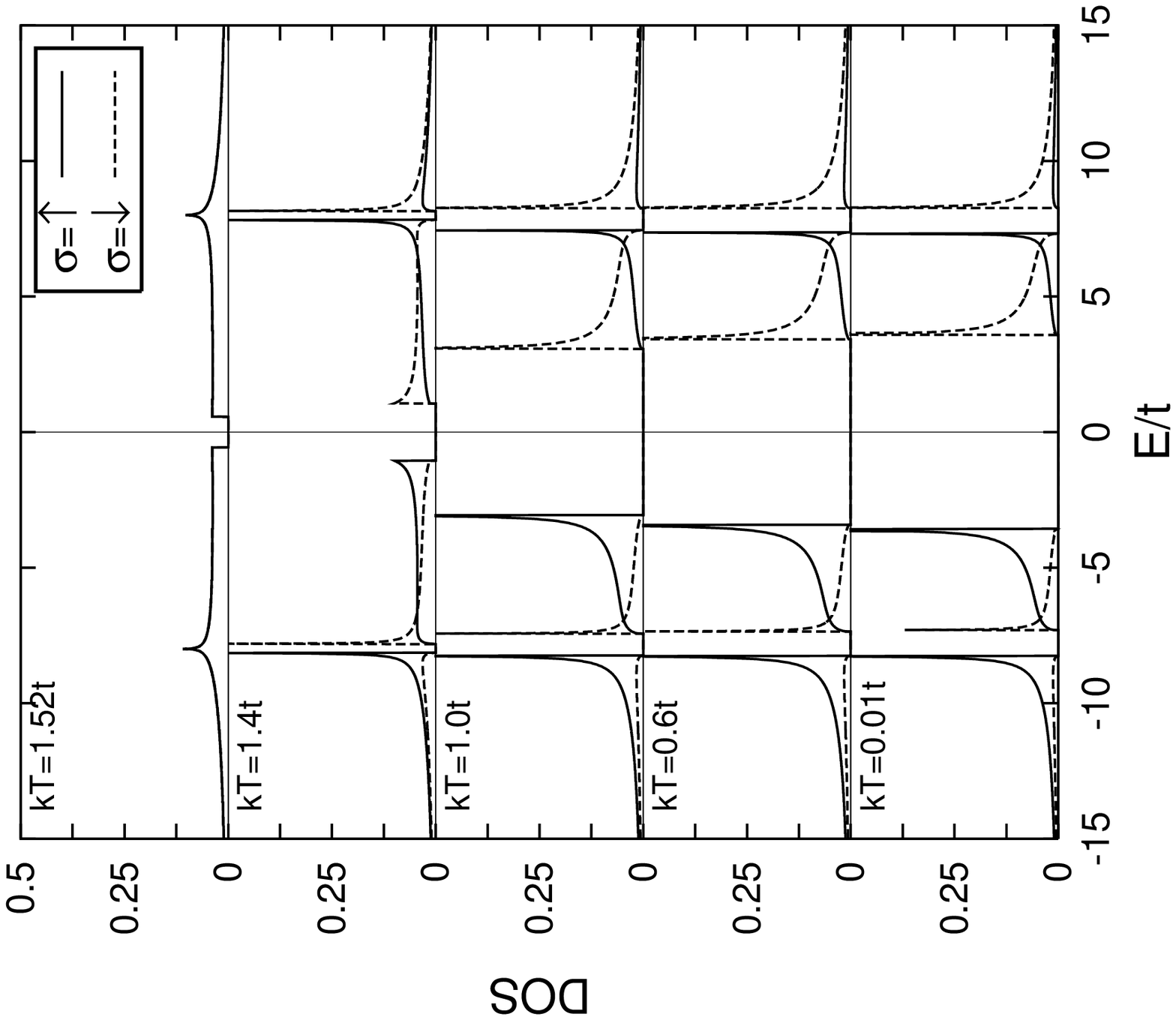}}
  \quad
  \subfigure[Sublattice Density of States for $U=10t$]
  {\label{fig:8b}
    \includegraphics*[width=6.4cm,angle=270]{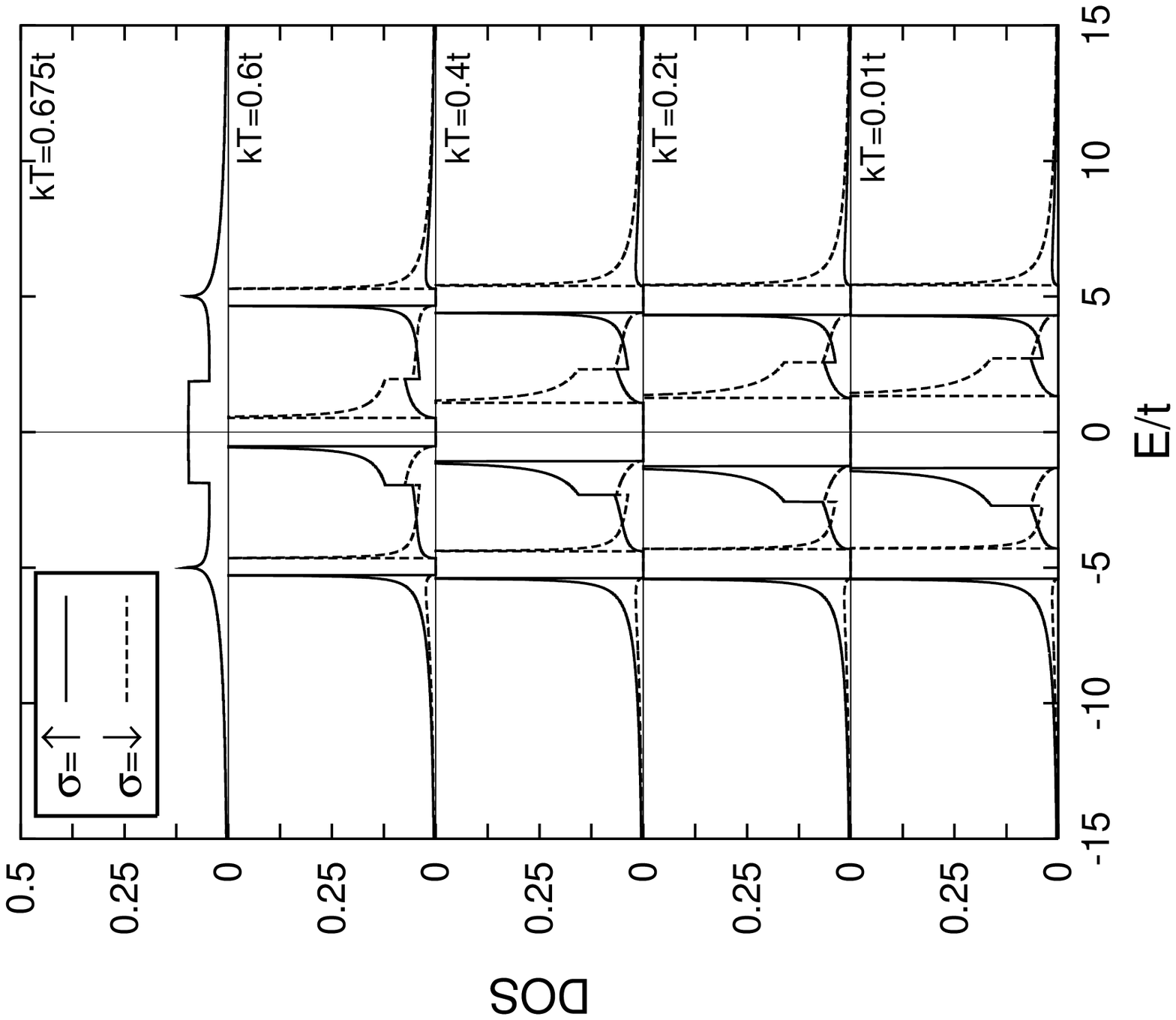}}
  \quad
  \subfigure[Band Structure on the Chemical Lattice for $U=10t$]
  {\label{fig:8c}
    \includegraphics*[width=6.4cm,angle=270]{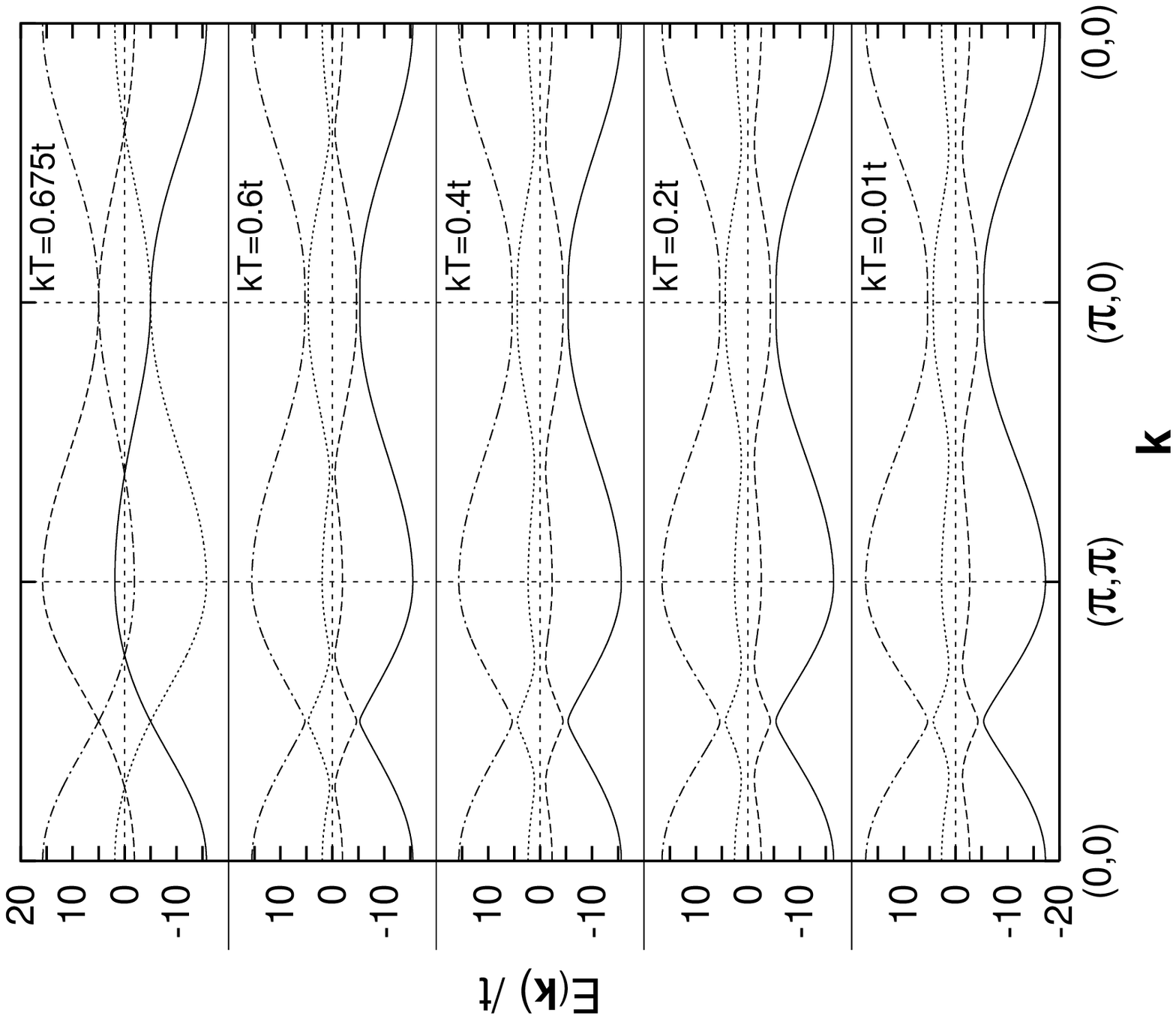}}
  \caption{Density of states and band structure at half filling; the thin
    vertical line at $E=0$ denotes the position of the Fermi level $E_F$.}
  \label{fig:8}
\end{figure}

The two-pole approximation leads to a splitting of the single electronic band
into two Hubbard subbands that correspond to the elementary excitations
described by the composite operators $\xi$ and $\eta$. Those two subbands are
separated by the on-site Coulomb interaction $U$, which may lead to a gap at
same critical value \cite{Georges:96,Mancini:00b}. The bipartite lattice
approach leads to a doubling of the two subbands by reflection around the band
center at $\frac U2 - \mu$ (see Fig.~\ref{fig:8c}). In the paramagnetic case,
when the magnetic lattice is introduced, this doubling of the Hubbard subbands
by the reduction of the Brillouin zone is completely artificial: the reflected
bands occupy exactly the same energy interval with the same spectral weights
as the original Hubbard subbands.

\begin{figure}[tbp]
  \centering
  \includegraphics*[width=5cm,angle=270]{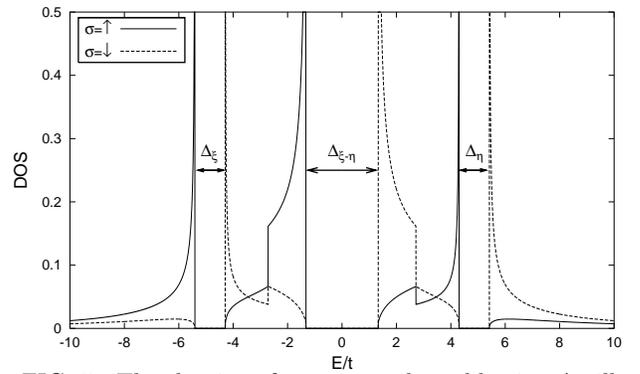}
  \caption{The density of states on the sublattice $\mathbf A$, illustrating
    the Mott-Heisenberg gaps in the lower and the upper Hubbard band as well
    as in their overlapping region in the paramagnetic
phase. The values of the parameters are: $U=10t$,
    $n=1.0$ and $kT=0.01t$.}
  \label{fig:7}
\end{figure}

The antiferromagnetic state, however, is characterized by the opening of
Mott-Heisenberg gaps at the crossing points of the four subbands induced by
the switching to the magnetic lattice. Thus we find three Mott-Heisenberg
gaps, one in the lower Hubbard band, $\Delta_{\xi}$, one in the upper Hubbard
band, $\Delta_{\eta}$ and a central Mott-Heisenberg gap,
$\Delta_{\text{$\xi$--$\eta$}}$, in the region where the two Hubbard subbands
overlap in the paramagnetic phase (see Fig.~\ref{fig:7}).

If the two Hubbard subbands are already separated by a Mott-Hubbard gap (in
the paramagnetic phase it has been found \cite{Mancini:00b} that the
Mott-Hubbard gap opens at $U\simeq13.2t$), the central Mott-Heisenberg gap
adds to this gap (see Fig.~\ref{fig:8a}). The gaps $\Delta_{\xi}$ and
$\Delta_{\eta}$ are not symmetric around the Mott separation $\pm U/2$,
because the upper and the lower band edges are shifted by different amounts.

The antiferromagnetic ordering thus leads to a splitting of the two Hubbard
subbands into four antiferromagnetic bands that are occupied by both, the
majority and the minority spin subsystems. As it can be seen from
Fig.~\ref{fig:7}, the two spin subsystems energetically occupy the same
regions -- which is the reason why the staggered magnetization does never
reach saturation $m=n$ --, but with rather different spectral weights: the
spectral weight of the majority spins is strongly enhanced at the upper band
edges, whereas the minority spins have an enhanced spectral weight at the
lower band edges, leading to the staggered magnetization. At half filling the
density of states shows a complete symmetry between the minority and majority
spins with respect to reflection of the energy around the Fermi level. This is
due to the fact that our solution respects the particle-hole symmetry.

\paragraph{The Mott-Heisenberg and the Mott-Hubbard gap}
\label{sec:4.2.2.1}

The interplay between the Mott-Hubbard gap and the Mott-Heisenberg gap at half
filling is illustrated in Fig.~\ref{fig:8}. For $U=10t$ the two Hubbard
subbands overlap at the N{\'e}el temperature $T_{\rm N}$ and by decreasing $T$
we find the opening of the Mott-Heisenberg gaps in the two Hubbard subbands as
well as in the central region (Fig.~\ref{fig:8b}). The corresponding evolution
of the electronic band structure in the Brillouin zone of the chemical lattice
is shown in Fig.~\ref{fig:8c}. We notice the doubling of the Hubbard subbands
in the paramagnetic state, which then evolves into the four antiferromagnetic
subbands by decreasing temperature, and the typical symmetry of the band
structure along the diagonal of the Brillouin zone for the chemical lattice
due to the reduced Brillouin zone of the magnetic lattice.

\end{multicols} \widetext
\begin{multicols}{2} \narrowtext

In Fig.~\ref{fig:8a} the case for $U=16t$ is shown, where the Mott-Hubbard gap
already separates the two Hubbard subbands in the paramagnetic phase just
above the transition point. By decreasing temperature the Mott-Heisenberg gaps
$\Delta_{\xi}$ and $\Delta_{\eta}$ open within each of the two Hubbard
subbands. In addition the central Mott-Heisenberg gap adds to the Mott-Hubbard
gap leading to the central gap $\Delta_{\text{$\xi$--$\eta$}}$.

\begin{figure}[tbp]
  \centering
  \subfigure[Central Gap $\Delta_{\text{$\xi$--$\eta$}}$ as Function of $T$]
  {\label{fig:9a}
    \includegraphics*[width=4.5cm,angle=270]{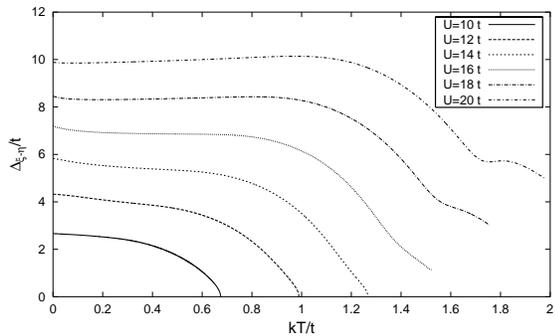}}
  \subfigure[Gaps in the Lower and the Upper Hubbard Band, $\Delta_{\xi}$ and
  $\Delta_{\eta}$, as Function of $T$]
  {\label{fig:9b}
    \includegraphics*[width=4.5cm,angle=270]{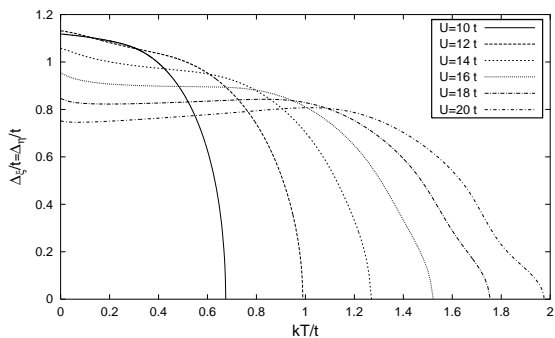}}
  \subfigure[Gaps in the Lower and the Upper Hubbard Band, $\Delta_{\xi}$ and
  $\Delta_{\eta}$, as Function of $U$]
  {\label{fig:9c}
      \includegraphics*[width=4.5cm,angle=270]{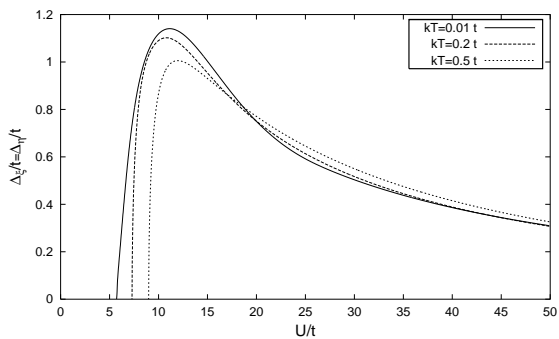}}
    \caption{The Mott and the Mott-Heisenberg gap at $n=1$.}
    \label{fig:9}
  \end{figure}

\begin{figure}[tbp]
  \centering
  \includegraphics*[width=8.0cm]{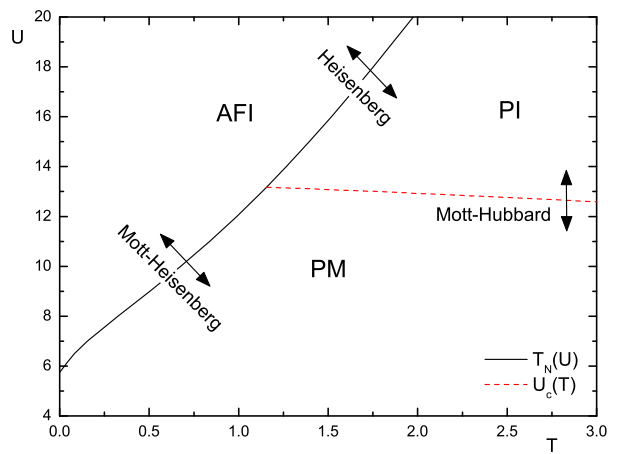}
  \caption{The variety of transitions at $n=1$.}
  \label{fig:9_1}
\end{figure}

The temperature dependence of these gaps is illustrated in Figs.~\ref{fig:9a}
and \ref{fig:9b} for various values of $U$. Above a certain critical value of
$U$ the central gap remains open at the phase transition. At low temperatures
the Mott-Heisenberg gaps within the two Hubbard subbands, $\Delta_{\xi}$ and
$\Delta_{\eta}$, are increasing as a function of $U$ up to a maximum value
$U\approx 11t$ and then are decreasing like $1/U$ (cf.\ Fig.~\ref{fig:9c}).
This reflects the $1/U$-dependence of the antiferromagnetic exchange integral
in the Heisenberg model, to which the half filled Hubbard model can be mapped
\cite{Anderson:63,Chao:77}. For the central gap the additional Mott separation
prevents this behavior. Also the N{\'e}el temperature does not show an $1/U$
like behavior (cf.\ Fig.~\ref{fig:9_1}), which indicates that the
antiferromagnetic exchange integral is not always properly taken into account.

\paragraph{The Metal Insulator Transition}
\label{sec:4.2.2.2}

The phenomenon of metal insulator transitions has been intensively studied in
the context of strongly correlated electron systems \cite{Imada:98}. In the
framework of the \emph{COM} the MIT within the paramagnetic state at half
filling is found\cite{Mancini:00b} in the light of the Hubbard picture, due to
separation of the two Hubbard subbands at a critical value of the Coulomb
interaction. This picture is quite different from the one found by the
\emph{DMFT}, where the MIT from a paramagnetic metal to a paramagnetic
insulator is mainly due to the vanishing of a narrow coherent quasi-particle
peak at the Fermi level as in the Gutzwiller approximation \cite{Georges:96}.
However, for an antiferromagnetic state on a bipartite lattice, this
interpretation can be maintained within the \emph{DMFT} only when frustration
by an additional $t^{\prime}$-hopping is introduced
\cite{Jarrell:93,Georges:96}.

At half filling, within the framework of the \emph{COM}, we find three kind of
transitions: a Mott-Heisenberg transition (i.e., a transition between a
paramagnetic metal and an antiferromagnetic insulator) mainly driven by the
temperature at low values of the Coulomb interaction; a Mott-Hubbard
transition (i.e., a transition between a paramagnetic metal and a paramagnetic
insulator), almost insensitive to the temperature, at high values of the
Coulomb interaction; an Heisenberg transition (i.e., a transition between an
antiferromagnet and a paramagnet), within the insulating phase, driven by the
temperature at high values of the Coulomb interaction. Moreover, the central
gap has two components depending on the value of the Coulomb interaction and
the temperature: one due to the antiferromagnetic correlations (in the
antiferromagnetic insulating phase) and another coming from the Mott-Hubbard
mechanism (in the paramagnetic and antiferromagnetic insulating phases). In
the Heisenberg transition the antiferromagnetic component of the central gap
vanishes (i.e., the magnetization disappears and the lateral gaps close up),
but the paramagnetic component remains finite. We have a finite critical value
of the Coulomb interaction for the Mott-Hubbard transition, in contrast to
what found, for instance, by the Hubbard I approximation and the \emph{SDA}.
This fact allowed us to study the Mott-Heisenberg transition existing at lower
value of the Coulomb interaction and that is obviously absent in any picture
based on the approximations mentioned above (they do not have the Mott-Hubbard
transition neither!). In Fig.~\ref{fig:9_1} we summarize the transitions
occurring at half filling within a treatment of the Hubbard model in the
framework of the \emph{COM}.

The filling controlled MIT is discontinuous at zero temperature. We have an
antiferromagnetic insulator at $n=1$ and an antiferromagnetic metal at $n<1$,
because the central gap is compensated by the discontinuity in the chemical
potential at $n=1$ and thus the Fermi level always lies inside the second
antiferromagnetic band for $n<1$. This is exactly the same result as found in
the usual mean-field approximation \cite{Imada:98} and in the \emph{DMFT}
without frustration \cite{Jarrell:93}.In the frustrated case with non-zero
$t^{\prime}$-hopping \emph{DMFT} as well as Quantum Monte Carlo studies
additionally led to a $U$ controlled MIT inside the antiferromagnetic phase at
$n=1$ \cite{Duffy:97c,Chitra:99}.

\begin{figure}[tbp]
  \centering
  \subfigure[$kT=0.2t$]
  {\label{fig:11a}\includegraphics*[width=4.7cm,angle=270]{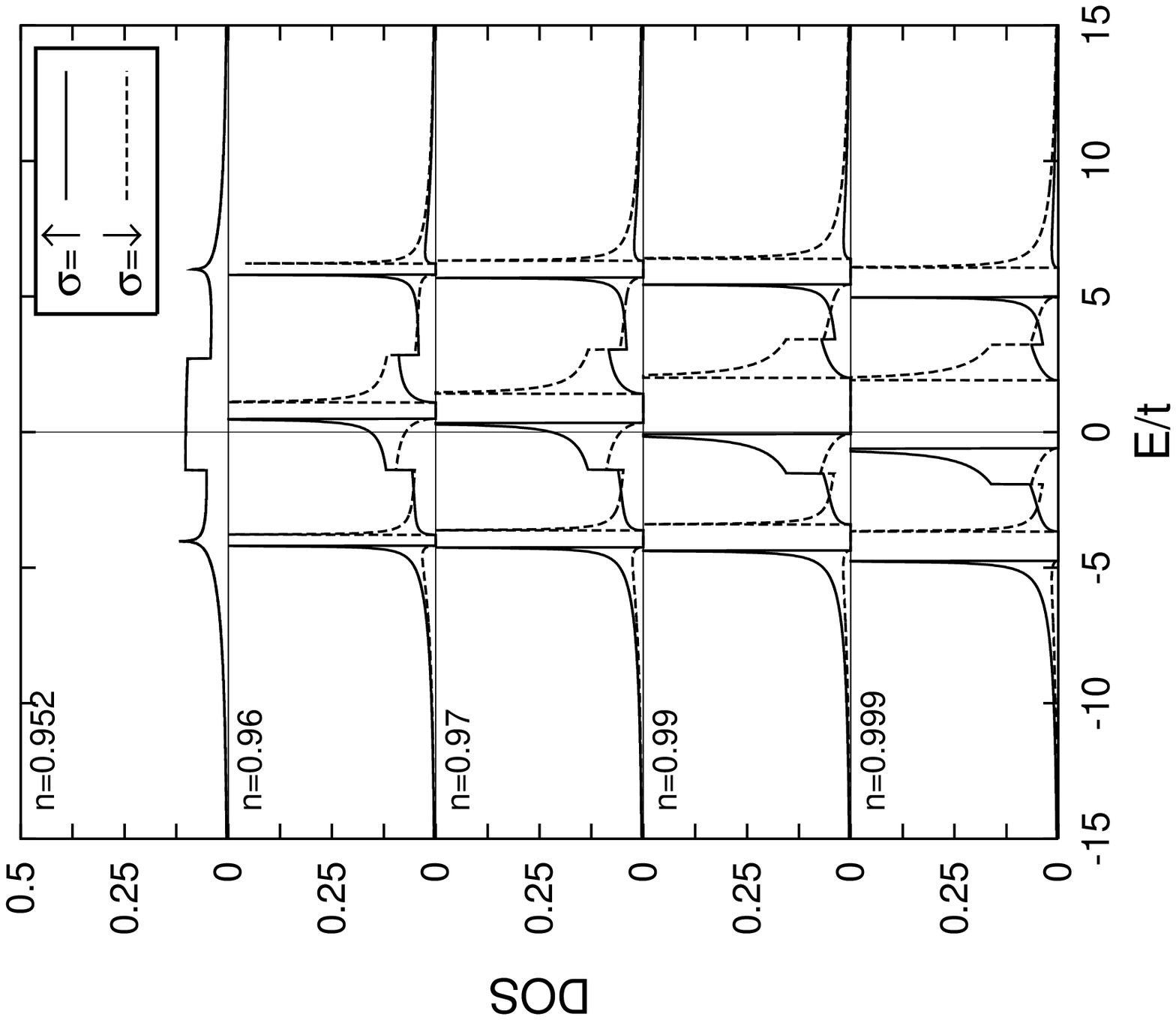}}
  \hfill
  \subfigure[$kT=0.5t$]
  {\label{fig:11b}\includegraphics*[width=4.7cm,angle=270]{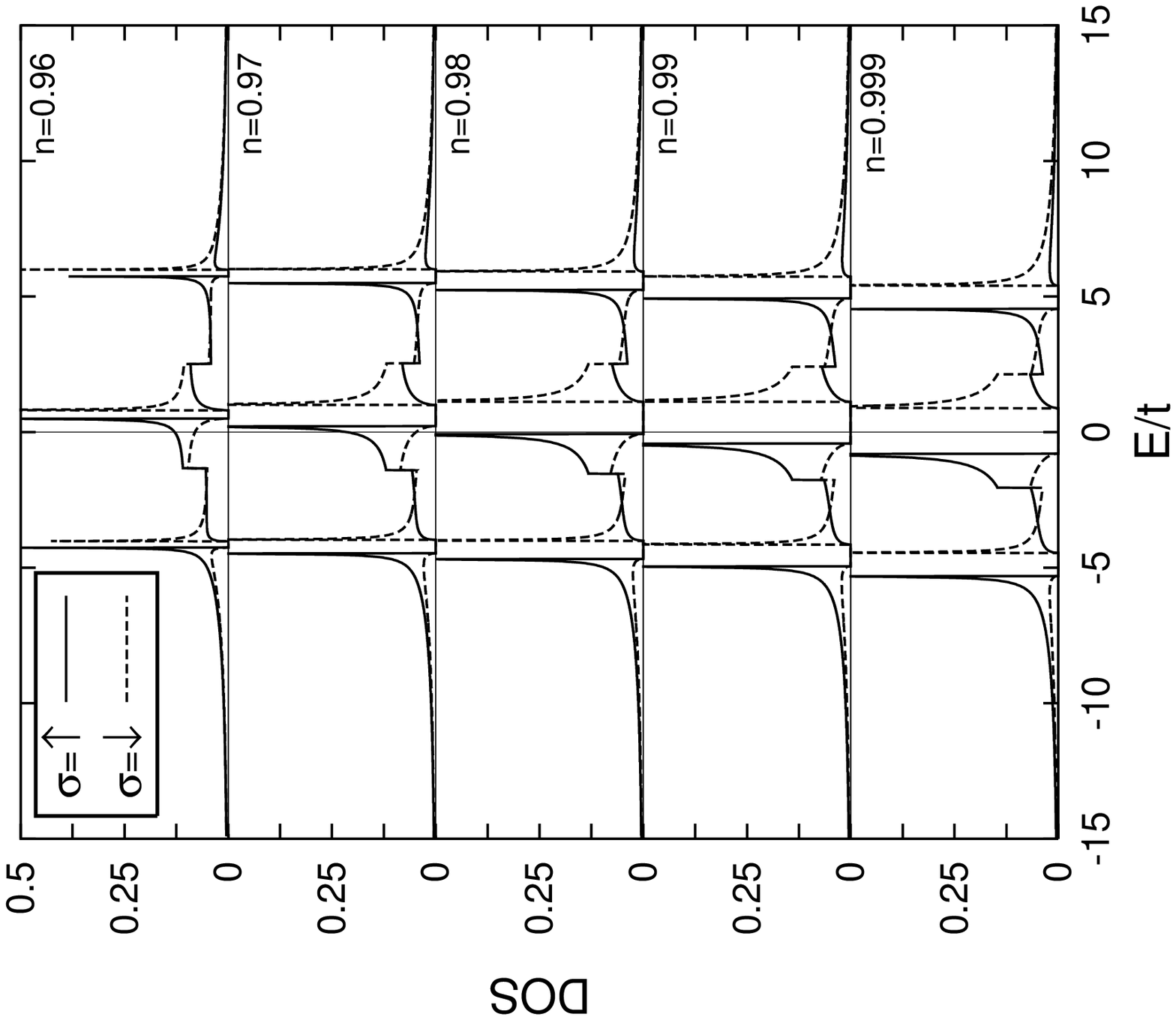}}
  \caption{The sublattice density of states at $U=10t$ for various values of
    the particle density. The thin vertical line at $E=0$ denotes the
    position of the Fermi level $E_F$.}
  \label{fig:11}
\end{figure}

Moving to finite temperature, the variation in the chemical potential, and
thus in the overall band shift, is considerably moderated for higher values of
$T$, whereas the central gap still remains large in comparison to $kT$.
Therefore, we find a finite region around half filling, where the Fermi level
is still situated inside $\Delta_{\text{$\xi$--$\eta$}}$ and we thus have an
antiferromagnetic phase of semiconductor type with very poor conductivity.
When the Fermi level is crossing the peak in the density of states at the
upper edge of the second antiferromagnetic band, a huge jump in the number of
carriers is observed and we finally get an antiferromagnetic metal (see
Fig.~\ref{fig:11}). Note that for the purely paramagnetic MIT such a filling
controlled transition is not to be expected.

Comparing Figs.~\ref{fig:11a} and \ref{fig:11b} we see how the moderate
evolution of the chemical potential as function of $n$ leads to a larger
extension of the insulating antiferromagnetic phase at higher temperatures.

The almost linear dependence of the central gap
$\Delta_{\text{$\xi$--$\eta$}}$ on the Coulomb interaction $U$ (cf.\
Fig.~\ref{fig:9a}) leads to a strong stabilization of the insulating
antiferromagnetic phase at $n=1$. Near half filling, however, this effect is
compensated for larger values of $U$ due to the strong decay of the chemical
potential when moving to lower particle densities. This decay of the chemical
potential then leads to a large overall shift of the band structure towards
higher energies, bringing the Fermi level close to the upper edge of the
second antiferromagnetic band. After a small region, where the insulating
phase is growing with $U$, this effects restricts the insulating
antiferromagnetic phase to regions very close to half filling for large values
of $U$.

\paragraph{The Shape of the $U$--$n$ Phase Diagram}
\label{sec:4.2.2.3}

\begin{figure}[tbp]
  \centering
  \includegraphics*[width=4.5cm,angle=270]{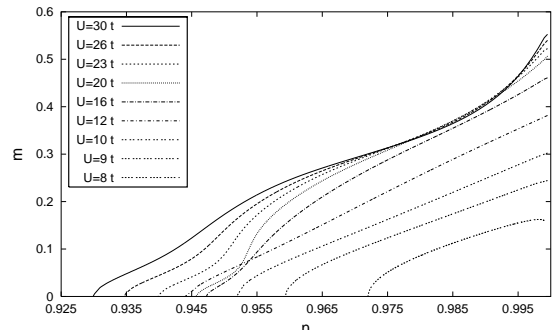}
  \caption{The sublattice magnetization as function of $n$ at $kT=0.2t$.}
  \label{fig:13}
\end{figure}

For large values of the Coulomb interaction close to the critical value $U_c$,
where antiferromagnetism is vanishing, the region of filling where the
antiferromagnetic phase exists is enlarged by increasing $U$. On the contrary,
for intermediate values of $U$ the antiferromagnetic region in doping shrinks.
This is can be explained by the closing of the Mott-Hubbard gap as a function
of doping. When at low values of $U$ no Mott-Hubbard gap is present, the
magnetization and the extension of the antiferromagnetic phase in $n$ grow
with growing $U$. After the opening of the Mott-Hubbard gap we find the system
in a region where the Mott-Hubbard gap closes in the proximity of $n=1$, i.e.,
already within the antiferromagnetic phase. This leads to a shift of the Fermi
level towards the middle of the second antiferromagnetic band where majority
and minority spins nearly have the same spectral weight and thus to a
considerable reduction of the staggered magnetization (cf.\
Fig.~\ref{fig:13}). For large values of $U$ the Mott-Hubbard gap remains open
within the whole region of doping, where the antiferromagnetic phase exists,
and we again find increasing stability of the antiferromagnetic phase with
increasing Coulomb interaction.

\paragraph{The Heat Magnetization}
\label{sec:4.2.2.4}

\begin{figure}[tbp]
  \centering
  \includegraphics*[width=4.5cm,angle=270]{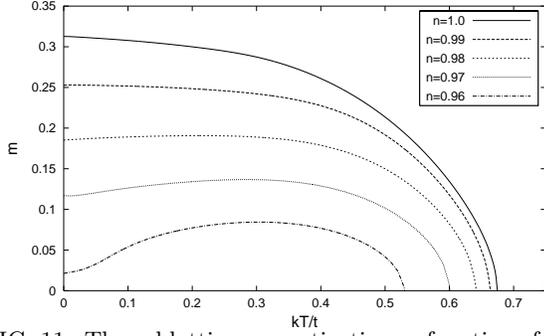}
    \caption{The sublattice magnetization as function of $T$ and $n$ at $U=10t$.}
    \label{fig:15}
  \end{figure}

\begin{figure}[tbp]
  \centering
  \includegraphics*[width=6.4cm,angle=270]{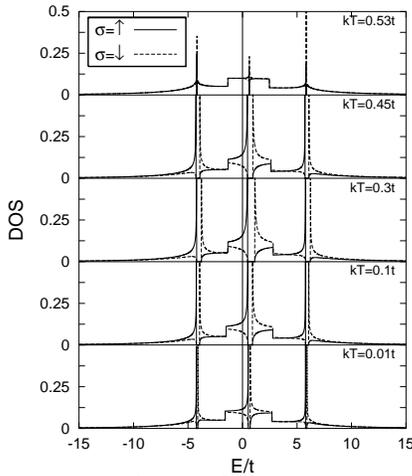}
  \caption{The sublattice density of states for $U=10t$ and $n=0.96$.}
  \label{fig:14}
\end{figure}

For small values of $U$ we find an increasing of the stability in $n$ of the
antiferromagnetic phase by increasing temperature. In the same manner, when
$n$ is close to the phase boundary, the magnetization itself first increases
with increasing $T$ before going down to zero, as is shown in
Fig.~\ref{fig:15}. This so-called `heat magnetization' has its explanation in
a strong change in the spectral weights of the spin subsystems by increasing
temperature, whereas the Mott-Heisenberg gaps and the chemical potential are
only subject to very little changes (see Fig.~\ref{fig:14}). At higher values
of $U$, after the opening of the Mott-Hubbard gap, the stronger Coulomb
interaction considerably stabilizes the antiferromagnetic state at low
temperatures and excludes the effect of heat magnetization.

\subsection{The Antiferromagnetic State of the $\mathbf 3D$ Hubbard Model}
\label{sec:4.3}

\begin{figure}[tbp]
  \centering
  \subfigure[Physical Phase Diagram at $U=10t$]
  {\label{fig:16a}
    \includegraphics*[width=5.2cm,angle=270]{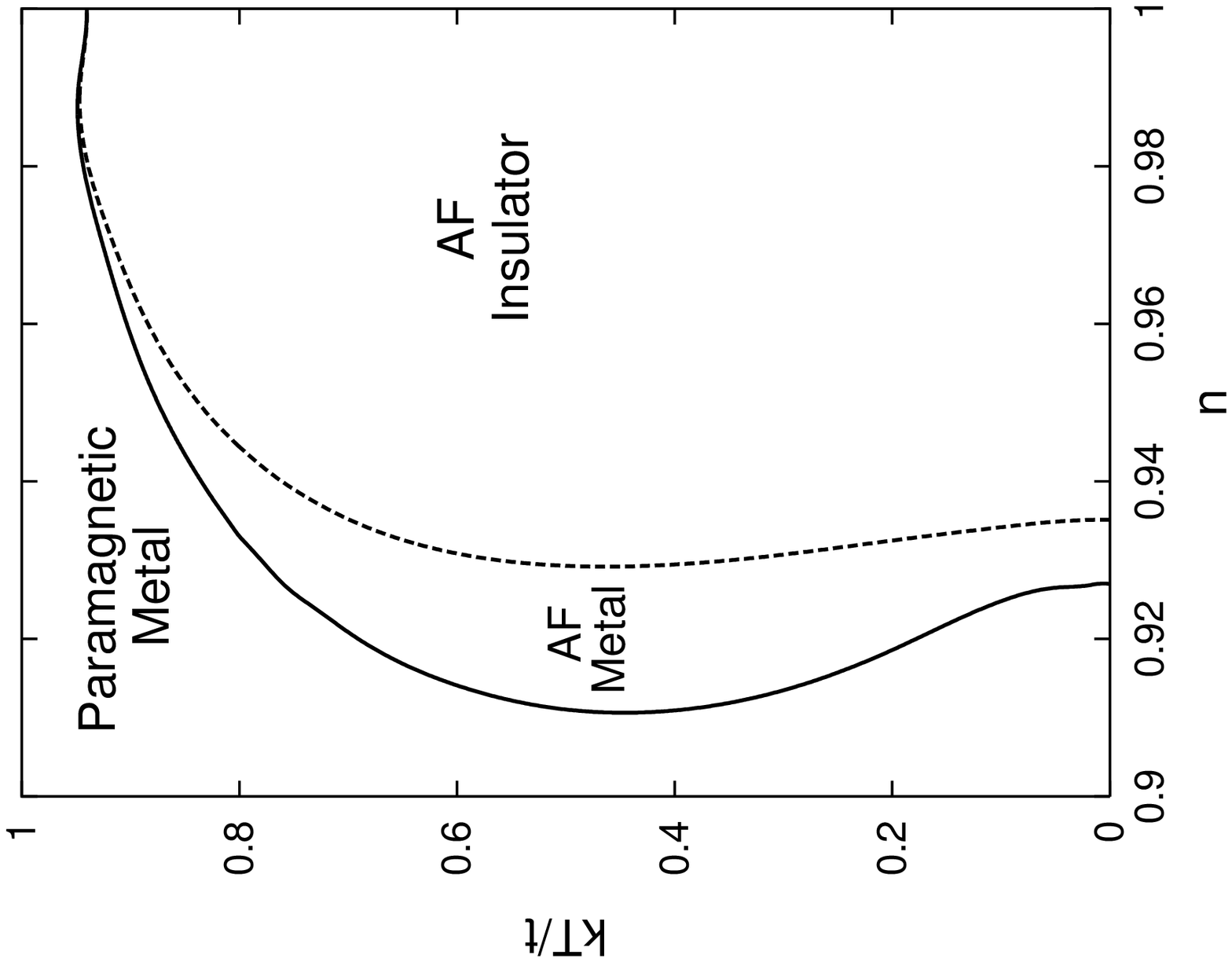}}
  \hskip 1mm
  \subfigure[Complete Phase Diagram at $U=23t$]
  {\label{fig:16c}
    \includegraphics*[width=5.2cm,angle=270]{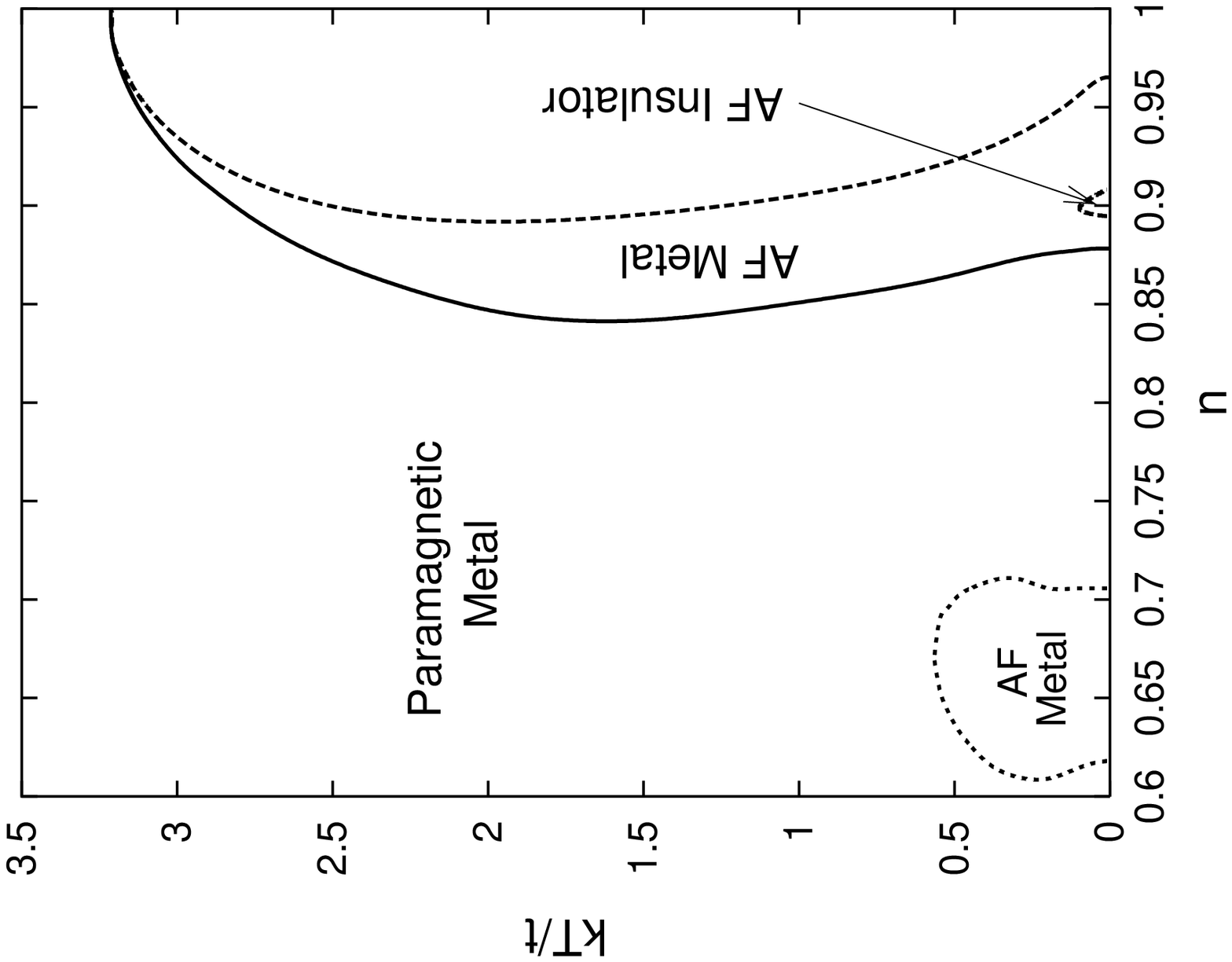}}
  \caption{The $n$--$T$ phase diagram for the 3D Hubbard model.}
  \label{fig:16}
\end{figure}

\begin{figure}[tbp]
  \centering
  \subfigure[Physical Phase Diagram at $kT=0.01t$]
  {\label{fig:17a}
    \includegraphics*[width=5.2cm,angle=270]{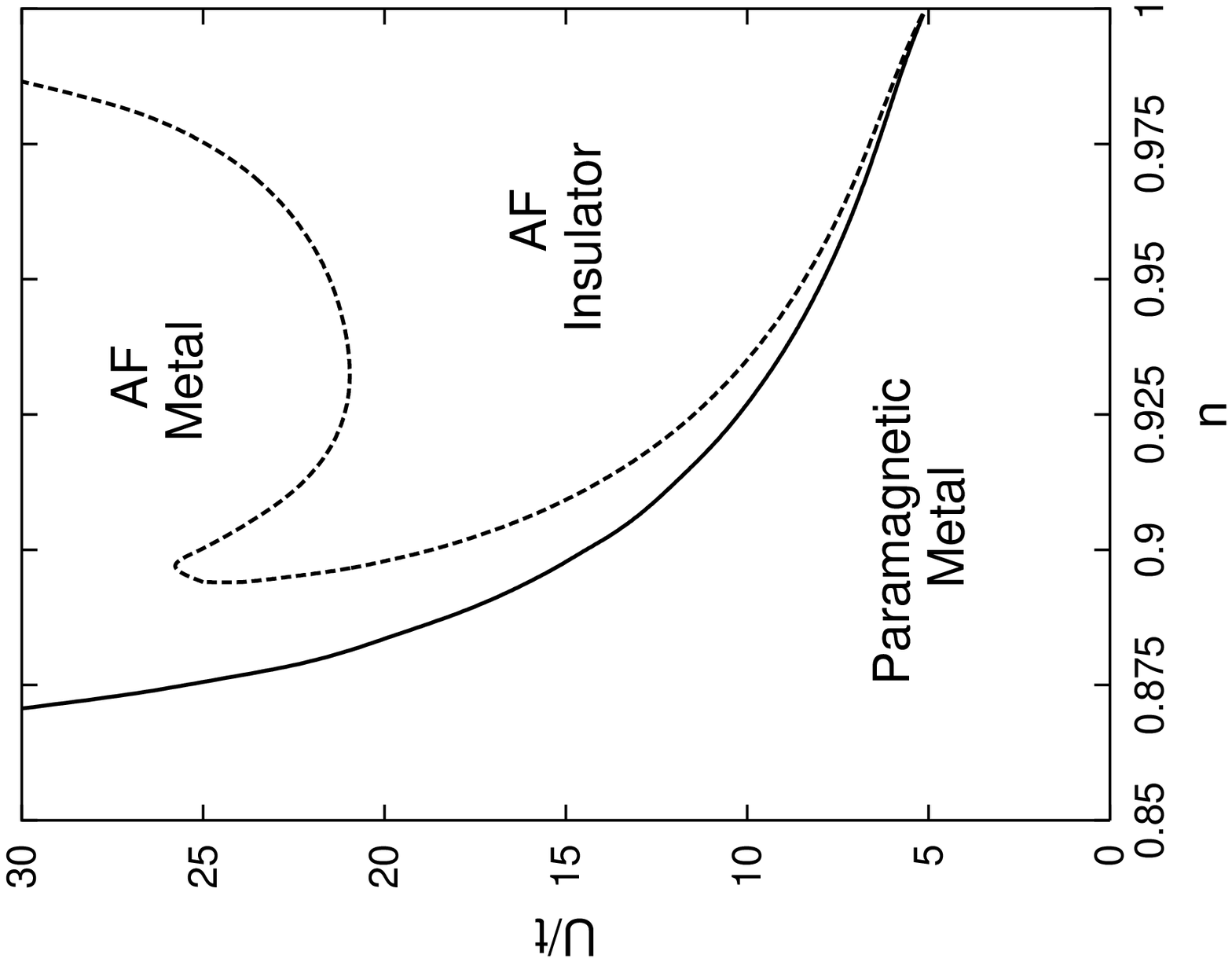}}
  \hskip 1mm
  \subfigure[Complete Phase Diagram at $kT=0.5t$]
  {\label{fig:17c}
    \includegraphics*[width=5.2cm,angle=270]{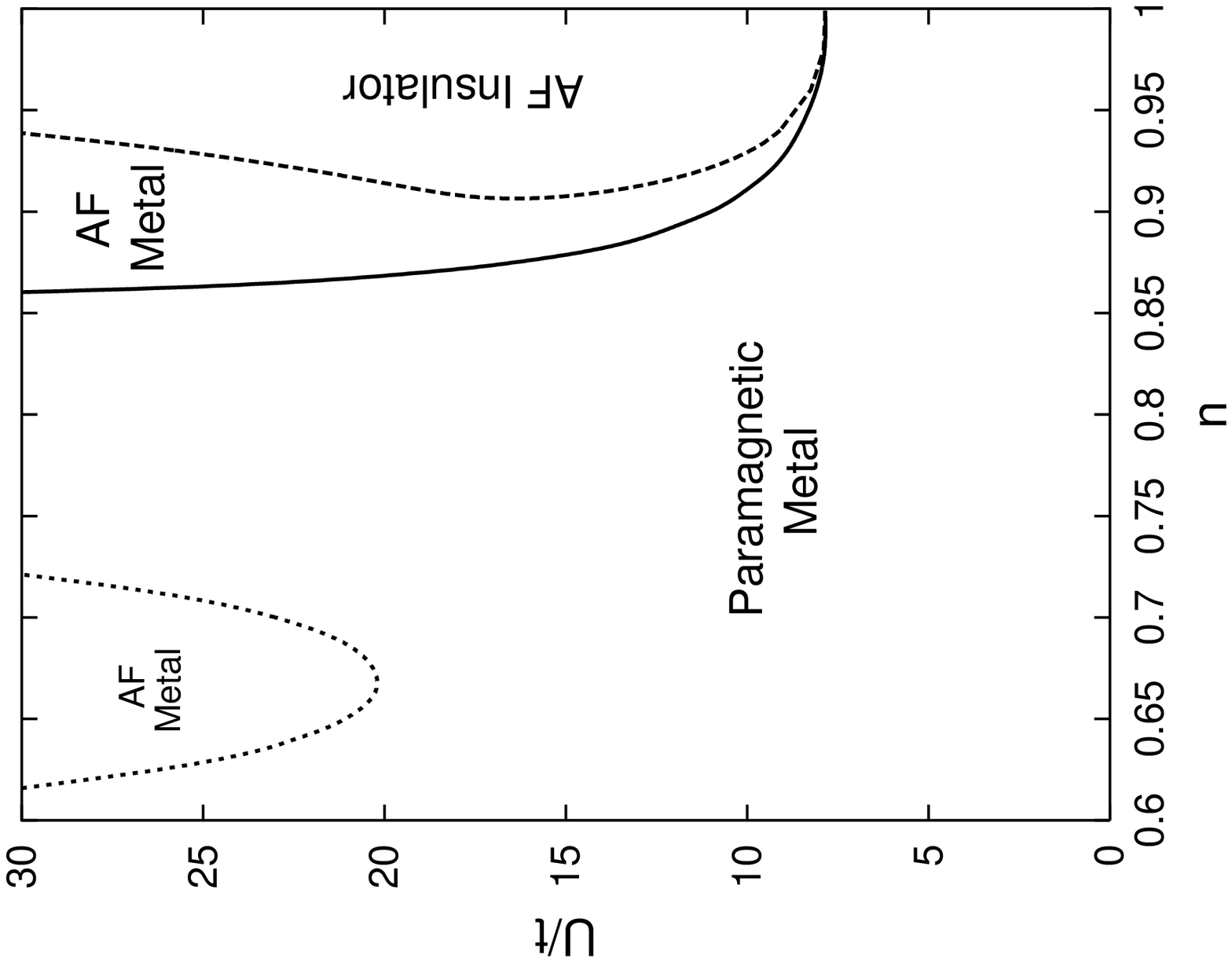}}
  \caption{The $U$--$n$ phase diagram for the 3D Hubbard model.}
  \label{fig:17}
\end{figure}

The antiferromagnetic phase of the $3D$ Hubbard model is very similar to the
one observed in $2D$. The shape and main features of the phase diagrams
(Figs.~\ref{fig:16} and \ref{fig:17}) remain unchanged (i.e., the MIT within
the antiferromagnetic phase, the finite critical Coulomb interaction for the
vanishing of the antiferromagnetic phase at half filling, the favoring of the
insulating state at high temperatures and low values of the Coulomb
interaction and the heat magnetization). Furthermore, as in the $2D$ case, all
these properties can be explained by analyzing the density of states and the
energy spectra.

The higher coordination number of the $3D$ model reduces the fluctuations and
leads to a greater stability of the antiferromagnetic phase as a function of
the external parameters $T$, $n$ and $U$. Anyway, comparing the extension in
$n$ of our antiferromagnetic phase with the one found within the \emph{SDA}
for a fcc-lattice \cite{Herrmann:97a} the former is restricted to a much
smaller region in doping.


The greater stability of the antiferromagnetic phase shows, within the range
of physically relevant values of the parameter $U$, that the metallic
antiferromagnetic phase extends to much lower values of $n$ (cf.\
Fig.~\ref{fig:16c} and \ref{fig:17c}) than it might be expected from the phase
diagrams in Figs.~\ref{fig:16a} and \ref{fig:17a}. For higher values of the
Coulomb interaction, up to $U=30t$, this additional antiferromagnetic region
is separated by the antiferromagnetic phase near half filling by a paramagnetic
region.

\begin{figure}[tbp]
  \centering
  \includegraphics*[width=4.5cm,angle=270]{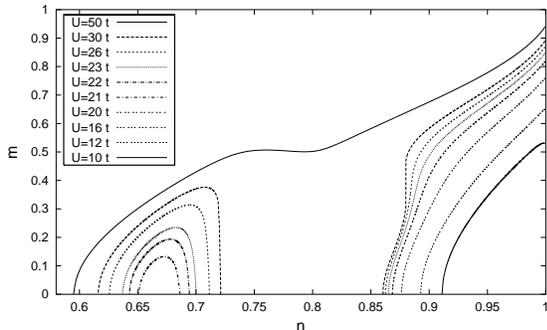}
  \caption{The sublattice magnetization as function of $n$ at $kT=0.5t$ for different
    values of the Coulomb interaction.}
  \label{fig:18}
\end{figure}

The form of the sublattice magnetization as a function of $n$ for different
values of the Coulomb interaction is given in Fig.~\ref{fig:18}. This result
indicates that the two antiferromagnetic regions in the complete phase
diagrams of Figs.~\ref{fig:16} and \ref{fig:17} are actually two parts of a
single antiferromagnetic solution, which are only connected at rather high
values of $U$. We will discuss in more details the nature of this `tail' in
the next section, in the context of the analysis of the extended Hubbard
model. Here we only remark that the separation of the two antiferromagnetic
phase regions by a paramagnetic one is due to the closing of the Mott-Hubbard
gap near $n=1$. This leads to a strong suppression of magnetization, as
described in Sec.~\ref{sec:4.2.2}\ref{sec:4.2.2.3} for the $2D$ case. The
`antiferromagnetic tail', which can be considered as an artifact of the
employed approximation, is also present in the $2D$ case, but only appears at
values of the Coulomb interaction $U\approx 100t$.

As we already mentioned in Sec.\ref{sec:4.2.2}, this `tail' at lower particle
densities could be energetically ruled out by other phases with magnetic or
spatial ordering.

\begin{figure}[tbp]
  \centering
  \subfigure[$n=0.9$]
  {\label{fig:19a}\includegraphics*[width=4.7cm,angle=270]{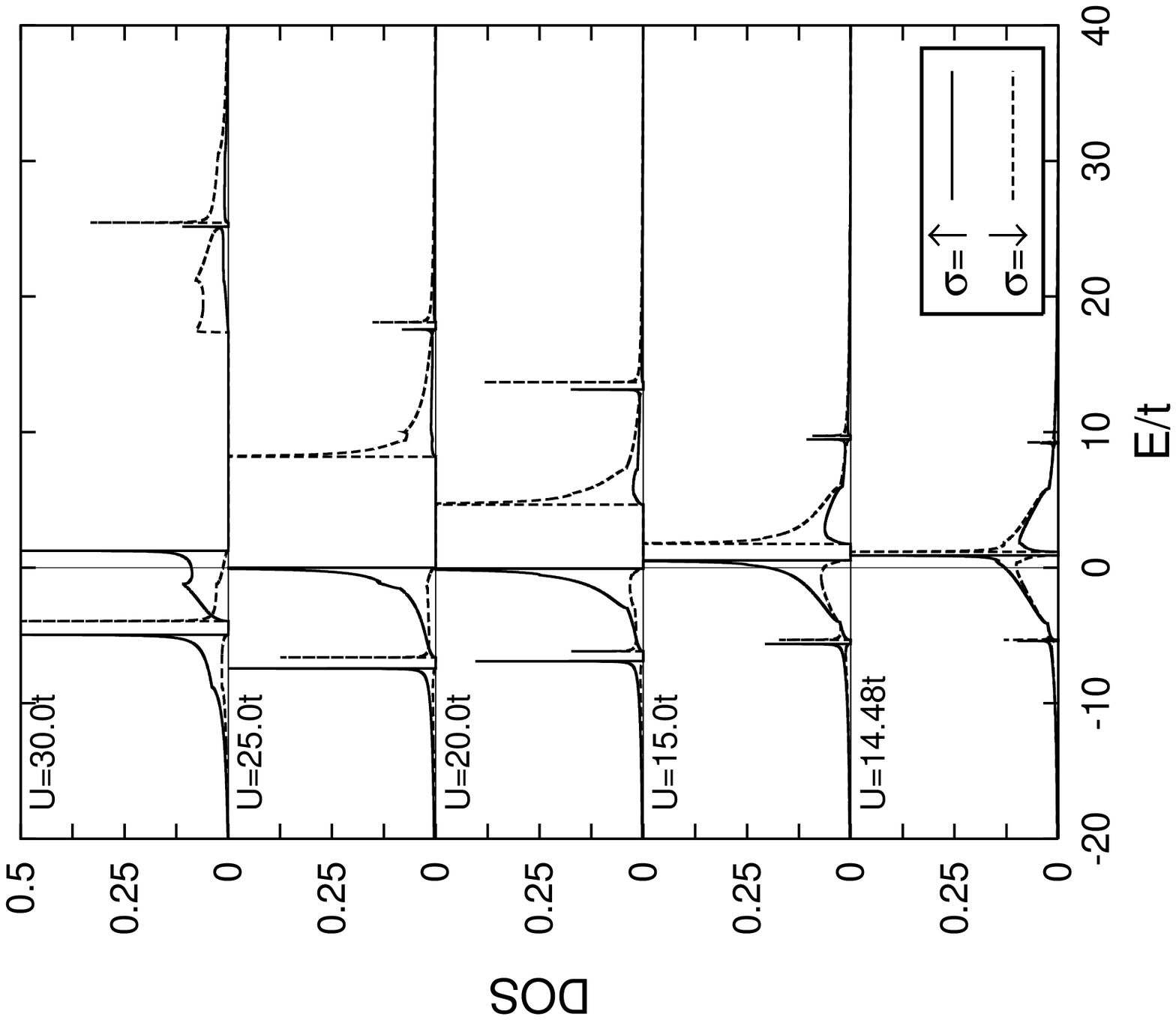}}
  \hfill
  \subfigure[$n=1.0$]
  {\label{fig:19b}\includegraphics*[width=4.7cm,angle=270]{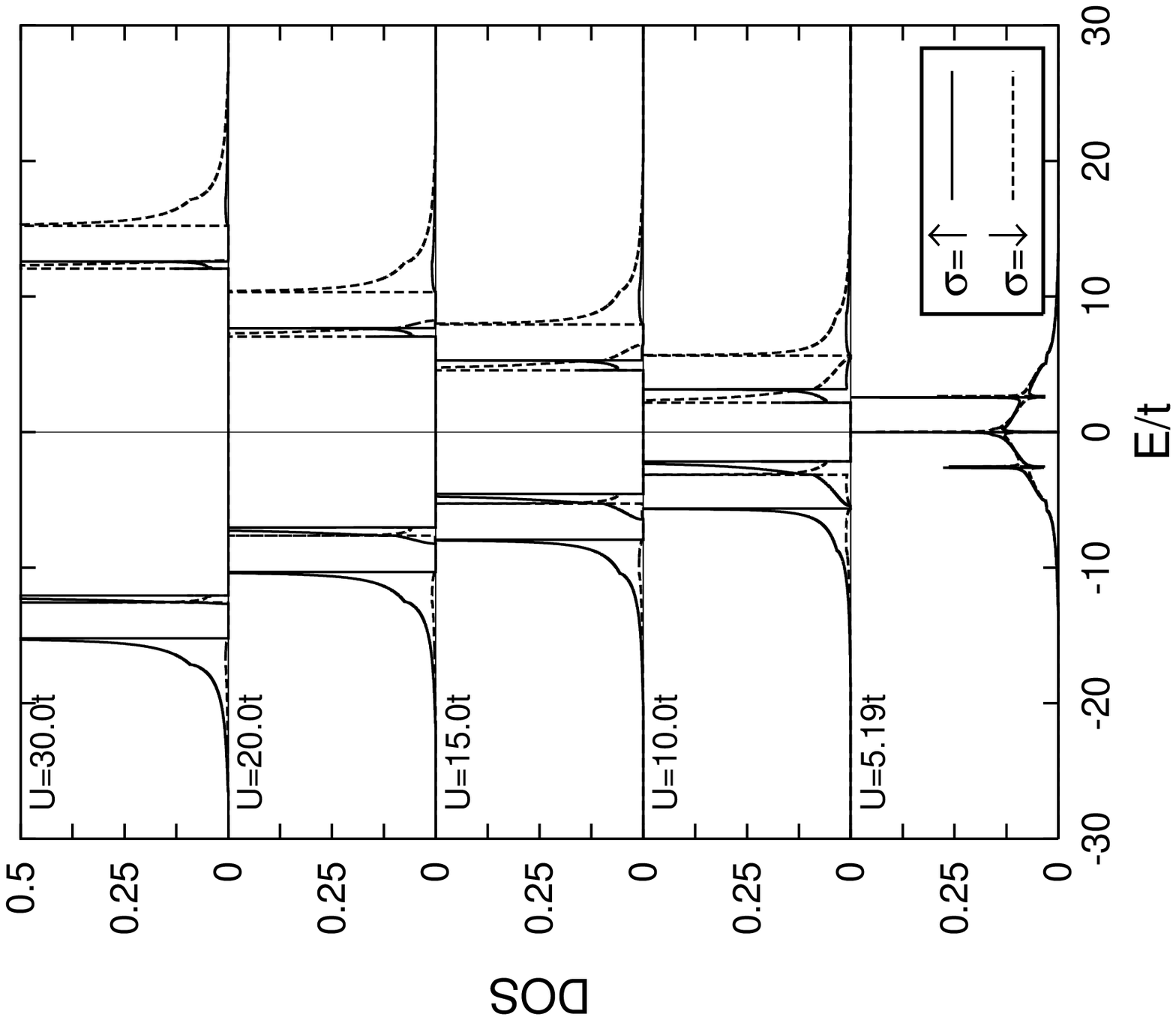}}
  \caption{The evolution of the sublattice density of states under variation
    of the Coulomb interaction at $kT=0.01t$. The thin vertical line at $E=0$ denotes
    the position of the Fermi level $E_F$.}
  \label{fig:19}
\end{figure}

Finally, in Fig.~\ref{fig:19}, we show the evolution of the sublattice density
of states by changing the Coulomb interaction. We note a broader structure in
comparison to the $2D$ model. As shown in Fig.~\ref{fig:19a}, the transitions
from metallic to semiconductor type antiferromagnetic states are related to the
interplay between the global band shift and the evolution of the central gap.
This explains the form of the insulating antiferromagnetic phase in
Fig.~\ref{fig:16a}.

\section{The Extended Hubbard Model}
\label{sec:5}

\subsection{The Model}
\label{sec:5.1}

In this section we study the antiferromagnetic solution of the \emph{extended
Hubbard model} in two dimensions. The model is described by the following
Hamiltonian
\begin{equation}
  \label{eq:5.1}
  \begin{split}
    H^{\text{\em ext}}&=\sum\limits_{
      \begin{subarray}{c}
        ij \\
        \sigma
      \end{subarray}
      } \left( t_{ij}-\mu \delta _{ij}\right)
    c_\sigma ^{\dagger }\left( i\right) c_\sigma \left( j\right)
    +U\sum\limits_in_{\uparrow }\left( i\right) n_{\downarrow }\left( i\right)\\
    &\qquad +\sum\limits_{
      \begin{subarray}{c}
        ij \\
        \sigma \sigma ^{\prime}
      \end{subarray}
      } V_{ij}n_\sigma
    \left( i\right) n_{\sigma^{\prime}}\left( j\right)\\[-4ex]
  \end{split}
\end{equation}
with the nearest neighbor Coulomb interaction
\begin{equation}
  \label{eq:5.2}
  V_{ij}=2V\alpha _{ij}=2V\frac 1N\sum_{{\mathbf k}}e^{i{\mathbf k\cdot }
    \left( {\mathbf R}_i-{\mathbf R}_j\right) }\alpha \left( {\mathbf
    k}\right) \,.
\end{equation}
We proceed along the guidelines given in Sec.~\ref{sec:3} to obtain the
antiferromagnetic state. The equation of motion for the fundamental
spinor~\eqref{eq:2.2} is given by

\end{multicols}

\widetext

\begin{equation}
  \label{eq:5.3}
  i\frac \partial {\partial t}\Psi \left( i,t\right) =\left[
    \Psi \left( i,t\right) ,H^{\text{\em ext}}\right]
  =\left(
    \begin{array}{l}
      -\mu \xi _{\uparrow }\left( i,t\right) -4tc_{\uparrow }^\alpha \left(
        i,t\right) -4t\pi _{\uparrow }^\alpha \left( i,t\right) +4Vn^\alpha \left(
        i,t\right) \xi _{\uparrow }\left( i,t\right) \\
      -\mu \eta _{\uparrow }\left( i,t\right) +U\eta _{\uparrow }\left( i,t\right)
      +4t\pi _{\uparrow }^\alpha \left( i,t\right) +4Vn^\alpha \left( i,t\right) \eta
      _{\uparrow }\left( i,t\right)  \\
      -\mu \xi _{\downarrow }\left( i,t\right) -4tc_{\downarrow }^\alpha \left(
        i,t\right) -4t\pi _{\downarrow }^\alpha \left( i,t\right) +4Vn^\alpha \left(
        i,t\right) \xi _{\downarrow }\left( i,t\right) \\
      -\mu \eta _{\downarrow }\left( i,t\right) +U\eta _{\downarrow }\left(
        i,t\right) +4t\pi _{\downarrow }^\alpha \left( i,t\right) +4Vn^\alpha \left(
        i,t\right) \eta _{\downarrow }\left( i,t\right)
    \end{array}
  \right) \, ,
\end{equation}
\begin{multicols}{2} \narrowtext \noindent where $n^\alpha \left( i\right) =\sum_{j\sigma }\alpha
_{ij}n_\sigma \left( j\right)$. The normalization matrix $I$ takes exactly the
form given in Eqs.~\eqref{eq:3.5} and \eqref{eq:3.7}. For the $m$-matrix we
get the same result as given in Eq.~\eqref{eq:3.8} without the terms coming
from the $t^{\prime}$-hopping, which are described by the projections
$\beta_{ij}$. The matrices $M_1,\ldots,M_4$ are built up from matrices
$\widehat{M}_1 ,\ldots ,\widehat{M}_4$, as in Eq.~\eqref{eq:3.9}, and these
latter have the following form
\end{multicols} \widetext
\begin{align*}
  \widehat{M}_1&=\left(
    \begin{array}{cc}
      -\mu \bigl( 1-\frac 12n\bigr) -4t\bigl( \Delta _{\downarrow }^\alpha
      +\Delta _{\uparrow }^\alpha \bigr) +4V\bigl( n-\upsilon _{\downarrow
        }-\upsilon _{\uparrow }\bigr)  & 4t\bigl( \Delta _{\downarrow }^\alpha
      +\Delta _{\uparrow }^\alpha \bigr)  \\[1ex]
      4t\bigl( \Delta _{\downarrow }^\alpha +\Delta _{\uparrow }^\alpha \bigr)  &
      \bigl( U-\mu \bigr) \frac 12n-4t\bigl( \Delta _{\downarrow }^\alpha +\Delta
      _{\uparrow }^\alpha \bigr) +4V\bigl( \upsilon _{\downarrow }+\upsilon
      _{\uparrow }\bigr)
    \end{array}
  \right)  \\[2ex]
  \widehat{M}_2&=\left(
    \begin{array}{cc}
      -\mu \frac m2 -4t\bigl( \Delta _{\downarrow }^\alpha -\Delta _{\uparrow }^\alpha
      \bigr) +4V\bigl( -\upsilon _{\downarrow }+\upsilon _{\uparrow }\bigr)  &
      4t\bigl( \Delta _{\downarrow }^\alpha -\Delta _{\uparrow }^\alpha \bigr)
      \\[1ex]
      4t\bigl( \Delta _{\downarrow }^\alpha -\Delta _{\uparrow }^\alpha \bigr)  &
      -\frac m2\bigl( -\mu +U\bigr) -4t\bigl( \Delta _{\downarrow }^\alpha -\Delta
      _{\uparrow }^\alpha \bigr) +4V\bigl( \upsilon _{\downarrow }-\upsilon
      _{\uparrow }\bigr)
    \end{array}
  \right)  \\[2ex]
  \widehat{M}_3&=\left(
    \begin{array}{cc}
      -4t\bigl( 1-n+p\bigr) +4V\lambda  & -4t\bigl( \frac 12n-p\bigr) +4V\bigl(
      \nu _{\uparrow }+\nu _{\downarrow }\bigr)  \\[1ex]
      -4t\bigl( \frac 12n-p\bigr) +4V\bigl( \nu _{\uparrow }+\nu _{\downarrow
        }\bigr)  & -4tp+4V\kappa
    \end{array}
  \right)  \label{eq:5.5}\tag{\theequation}\addtocounter{equation}{1} \\[2ex]
  \widehat{M}_4&=\left(
    \begin{array}{cc}
      0 & -2tm+4V\bigl( \nu _{\uparrow }-\nu _{\downarrow }\bigr)  \\[1ex]
      2tm-4V\bigr( \nu _{\uparrow }-\nu _{\downarrow }\bigr)  & 0
    \end{array}
  \right)\, ,
\end{align*}
\begin{multicols}{2} \narrowtext where we define the additional parameters
\begin{equation}
  \label{eq:5.6}
  \begin{split}
    \lambda  &=\bigl\langle \xi _\sigma \left( i\right) \xi _\sigma ^{\alpha
        \dagger }\left( i\right) \bigr\rangle =\bigl\langle \xi _{\overline{\sigma }
        }\left( i\right) \xi _{\overline{\sigma }}^{\alpha \dagger }\left( i\right)
    \bigr\rangle \quad ,\quad \text{for }i\in {\mathbf A} \\
    \nu _\sigma  &=\frac 12\bigl\langle \xi _\sigma \left( i\right) \eta
      _\sigma ^{\alpha \dagger }\left( i\right) \bigr\rangle \quad ,\quad
    \text{for }i\in {\mathbf A} \\
    \kappa  &=\bigl\langle \eta _\sigma \left( i\right) \eta _\sigma ^{\alpha
        \dagger }\left( i\right) \bigr\rangle =\bigl\langle \eta _{\overline{\sigma
          }}\left( i\right) \eta _{\overline{\sigma }}^{\alpha \dagger }\left(
        i\right) \bigr\rangle \quad ,\quad \text{for }i\in {\mathbf A} \\
    \upsilon _\sigma  &=\frac 12\bigl\langle n^{\alpha} \left( i\right)
      n_\sigma \left( i\right) \bigr\rangle \quad ,\quad
    \text{for }i\in {\mathbf A}\, .
  \end{split}
\end{equation}
We remark that the parameters $\lambda$ and $\kappa$ are not spin dependent
because of the requirement for the $m$-matrix to be real and the additional
symmetry constraint in Ass.~\ref{ass:1}.3. Furthermore, the parameters $\Delta
_\sigma^\alpha$ can be expressed through the parameters from
Eq.~\eqref{eq:5.6} as
\begin{equation}
\Delta _\sigma ^\alpha =\frac 12\lambda +\nu _\sigma -\nu
_{\overline{\sigma}}-\frac 12\kappa\,.
\end{equation}

The single-particle retarded Green's functions are calculated as in
Eqs.~\eqref{eq:3.21} and \eqref{eq:3.26}, just omitting the parts
corresponding to the $t^{\prime}$-hopping. The parameters in
Eq.~\eqref{eq:5.5} can then be calculated self-consistently by means of the
correlation functions~\eqref{eq:3.28}. The parameters $m$ and $\mu$ are
calculated as in Eq.~\eqref{eq:3.30a}. The parameters $\lambda$,
$\nu_{\sigma}$ and $\kappa$ are directly related to the single-particle
retarded Green's functions by
\begin{equation}
  \label{eq:5.7}
  \begin{split}
    \lambda &=C_{11}^{AB\widetilde{\alpha }
      }( \widetilde{{\mathbf R}}_i) \\
    \nu _{\uparrow }&=\frac 12C_{12}^{AB\widetilde{
        \alpha }}( \widetilde{{\mathbf R}}_i)\\
    \nu _{\downarrow }&=\frac 12C_{12}^{BA\widetilde{
        \alpha }}( \widetilde{{\mathbf R}}_i)\\
    \kappa &=C_{22}^{AB\widetilde{\alpha }}(
      \widetilde{{\mathbf R}}_i)
  \end{split}
\end{equation}
and the parameters $p$ and $\upsilon_{\sigma}$, which derive from higher-order
correlation functions are used as in Eq.~\eqref{eq:3.30c} to satisfy the
algebraic relations corresponding to the Pauli principle on the level of
thermal equilibrium states
\begin{align*}
  C_{12}^{AA}( \widetilde{{\mathbf R}}_i) &=0\\
  C_{12}^{BB}( \widetilde{{\mathbf R}}_i) &=0\\
  C_{11}^{AA}( \widetilde{{\mathbf R}}_i) &=C_{11}^{BB}(
    \widetilde{{\mathbf R}}_i)\, .
\end{align*}
Again, all the self-consistent equations are coupled.

\subsection{The Antiferromagnetic State of the $\mathbf 2D$ Extended Hubbard
  Model}
\label{sec:5.2}

\begin{figure}[tbp]
  \centering
  \subfigure[The $n$--$T$ Phase Diagram at $U=8t$]
  {\label{fig:20a}
    \includegraphics*[width=6.9cm,angle=270]{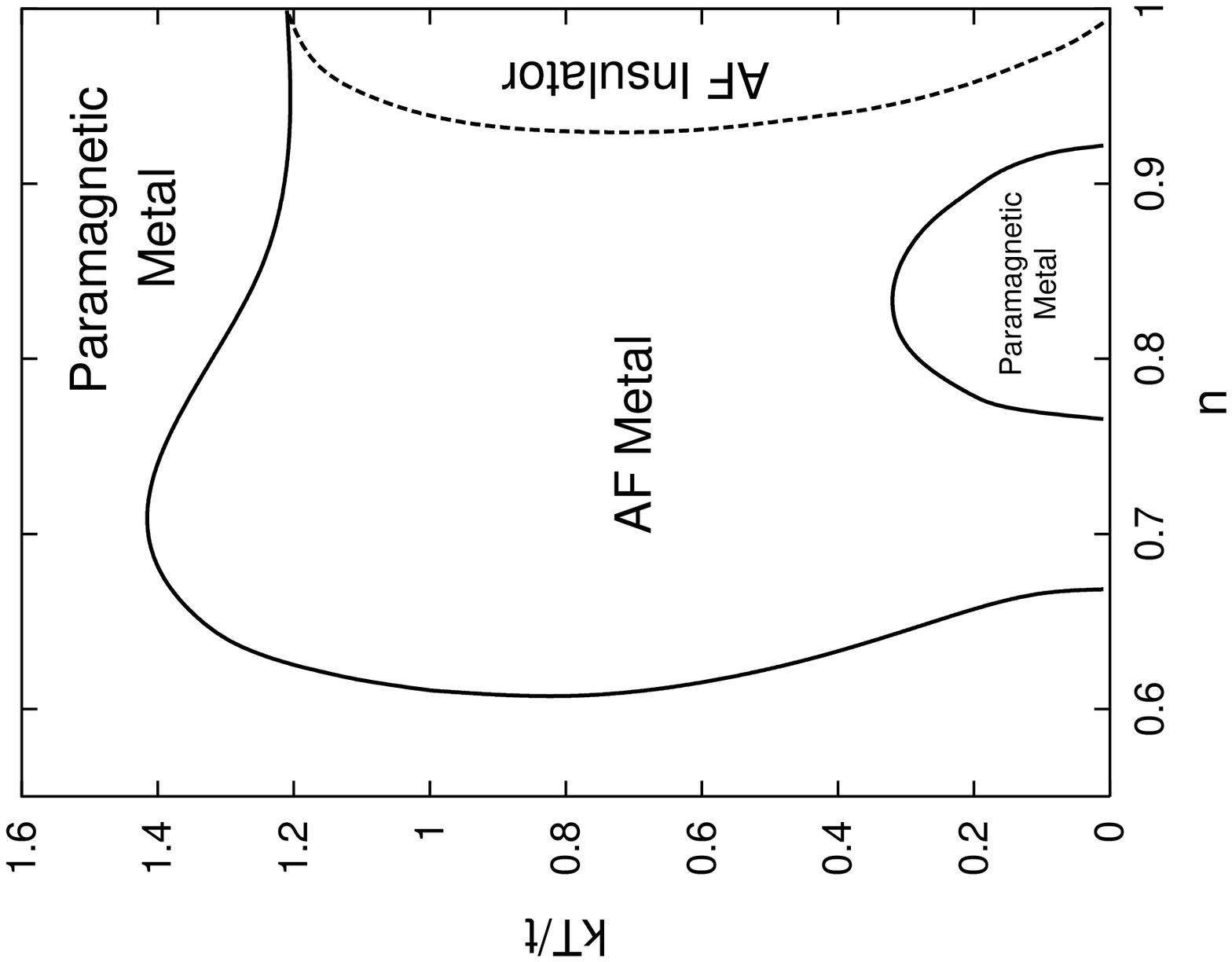}}
  \quad
  \subfigure[The $U$--$n$ Phase Diagram at $kT=0.01t$]
  {\label{fig:20b}
    \includegraphics*[width=6.9cm,angle=270]{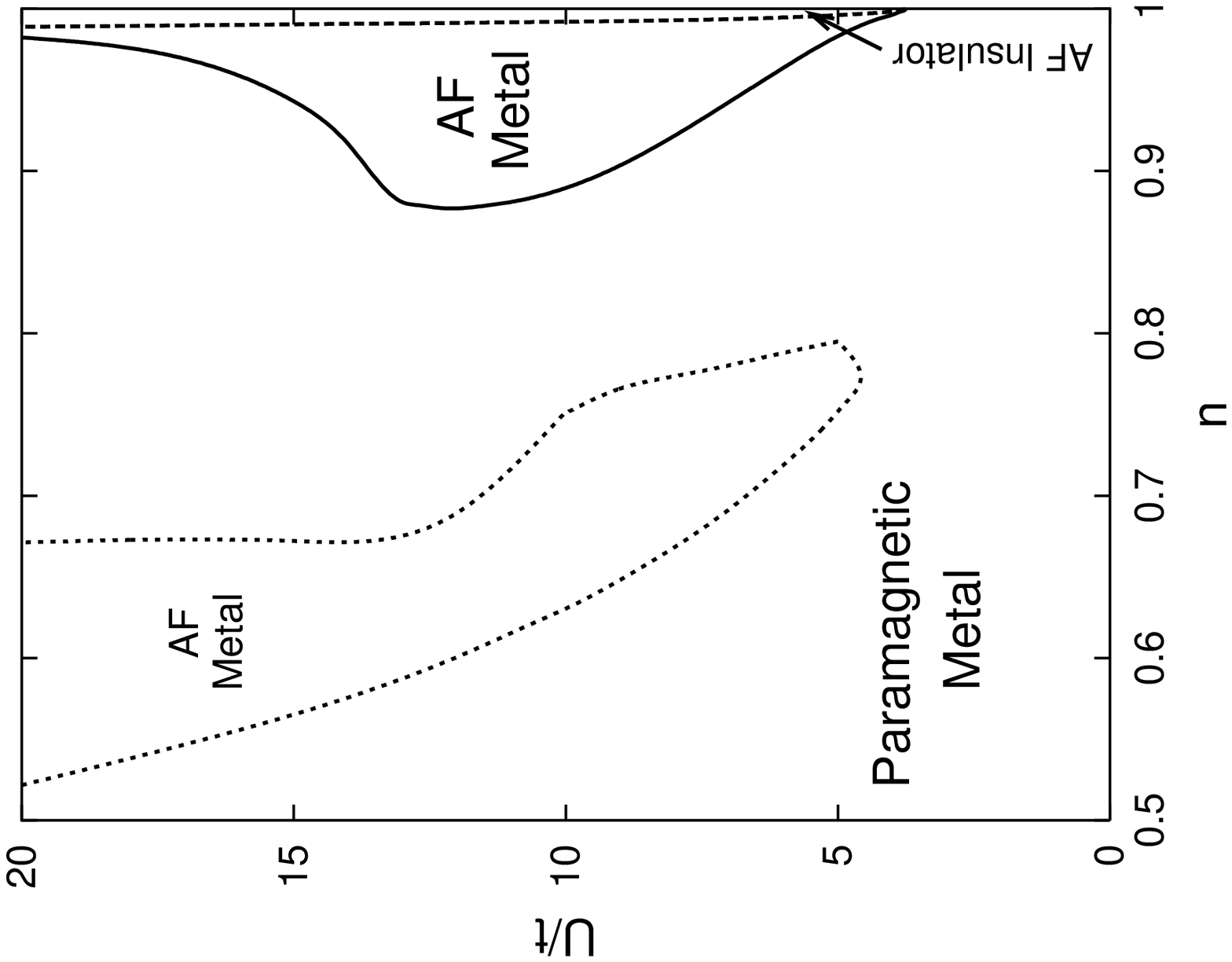}}
  \quad
  \subfigure[The $U$--$n$ Phase Diagram at $kT=0.4t$]
  {\label{fig:20c}
    \includegraphics*[width=6.9cm,angle=270]{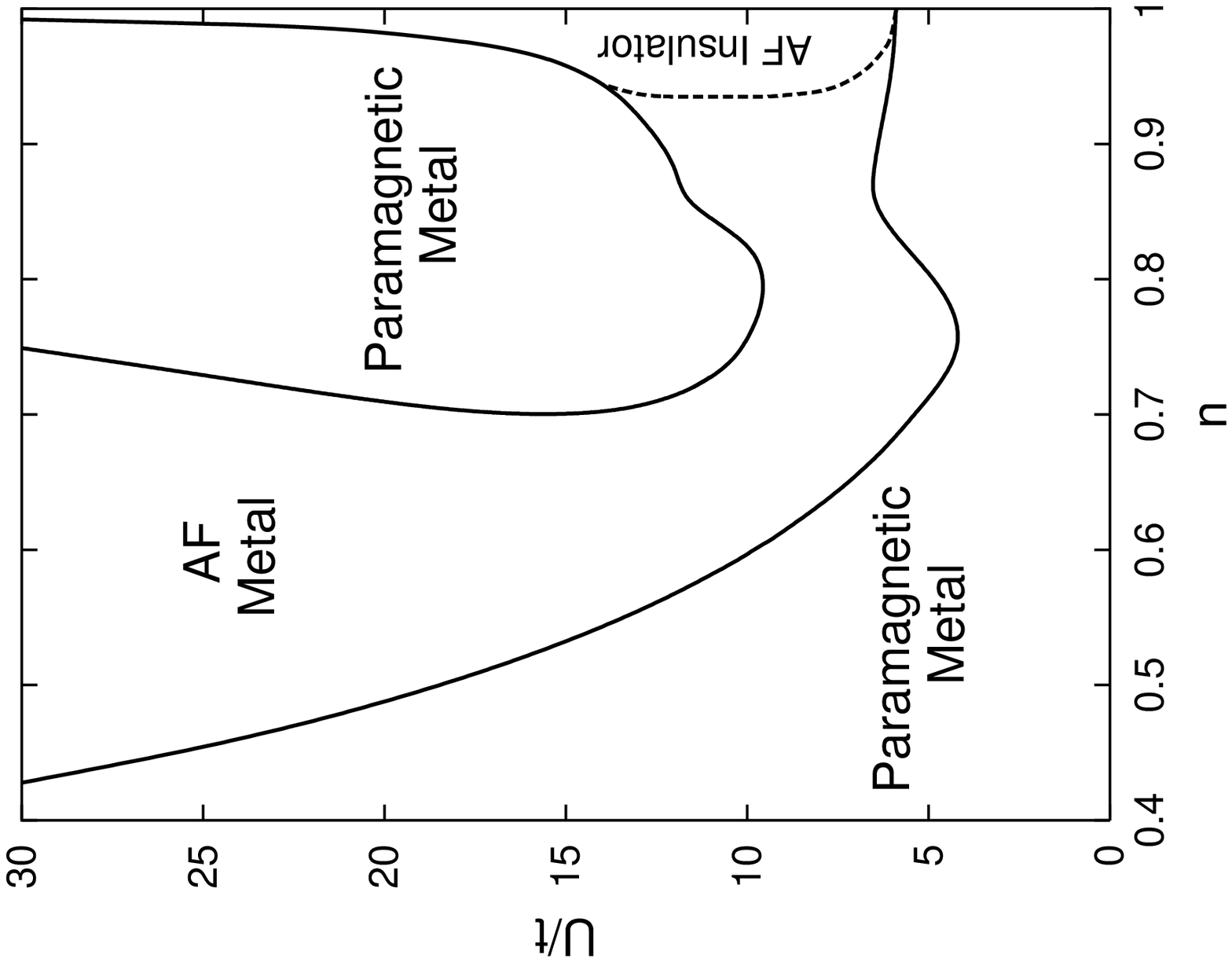}}
  \caption{The phase diagrams for the extended Hubbard model with $V=U/5$.}
  \label{fig:20}
\end{figure}

The phase diagram corresponding to the antiferromagnetic thermal equilibrium
state obtained as solution of the self-consistent equations~\eqref{eq:5.7},
\eqref{eq:3.30a} and \eqref{eq:3.30c} is shown in Fig.~\ref{fig:20} where we
took a fixed ratio $U/V=5$. Again, the antiferromagnetic phase always has a
lower free energy with respect to the paramagnetic phase and the phase
transition is always of second order.

The antiferromagnetic thermal equilibrium state fulfills the particle-hole
symmetry. In addition, the main qualitative features of the antiferromagnetic
states for the simple Hubbard model in two and three dimensions can be found in
the extended Hubbard model, too. There is a critical value of the Coulomb
interaction, for which the antiferromagnetic phase disappears and a metal
insulator transition occurs.

The phenomenon of heat magnetization is more pronounced than in the simple
Hubbard model and the occurrence of a `tail' in the antiferromagnetic solution
is observed down to rather low values of $U$, where they can even join for
higher temperatures. This leads to paramagnetic inclusions within the
antiferromagnetic phase. For higher values of $U$ the antiferromagnetic phase
is reduced to a very narrow region near half filling.

The band structure turns out to be somehow different from the one found for
the simple Hubbard model. The presence of the inter-site Coulomb interaction
leads to a large overlap of the antiferromagnetic bands, such that the opening
of the Mott-Heisenberg gap at the crossing points cannot split the lower and
the upper Hubbard band in most part of the antiferromagnetic phase region. The
sublattice density of states for the extended Hubbard model is thus mainly
characterized by a single central gap, which is the superposition of the
Mott-Heisenberg and the Mott-Hubbard gap (cf.\ Fig.~\ref{fig:21}).
\end{multicols} \widetext
\begin{multicols}{2} \narrowtext The part of the antiferromagnetic phase at lower particle density
is always characterized by this overlap of the antiferromagnetic subbands
emerging from the lower and the upper Hubbard band respectively. The absence
of the Mott-Heisenberg gap within the Hubbard subbands reduces the number of
peaks in the density of states to four:
\begin{figure}[tbp]
  \centering
  \subfigure[Sublattice Density of States for $U=9.5t$]
  {\label{fig:21a}\includegraphics*[width=4.7cm,angle=270]{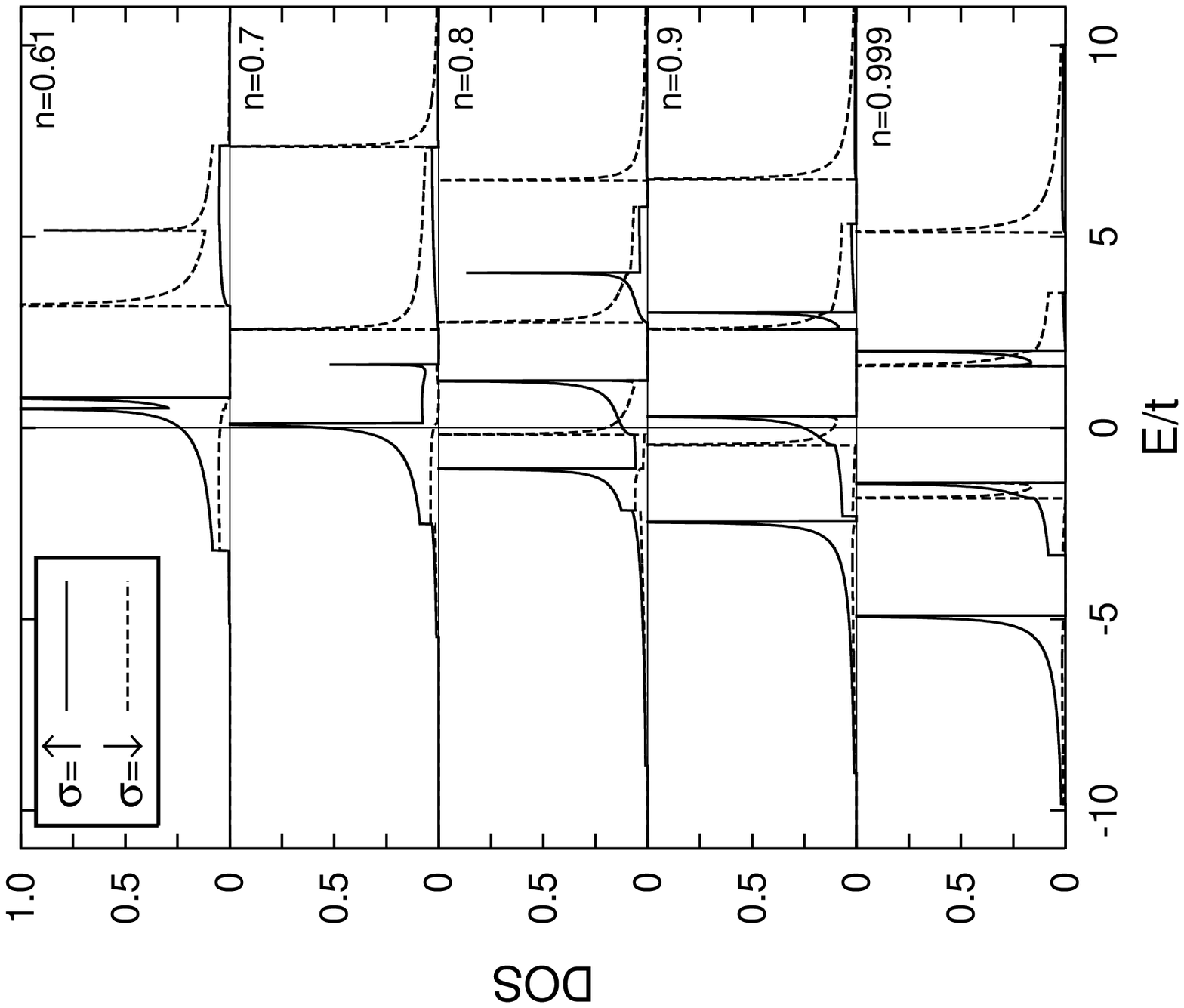}}
  \hfill
  \subfigure[Band Structure on the Chemical Lattice for $U=9.5t$]
  {\label{fig:21b}\includegraphics*[width=4.7cm,angle=270]{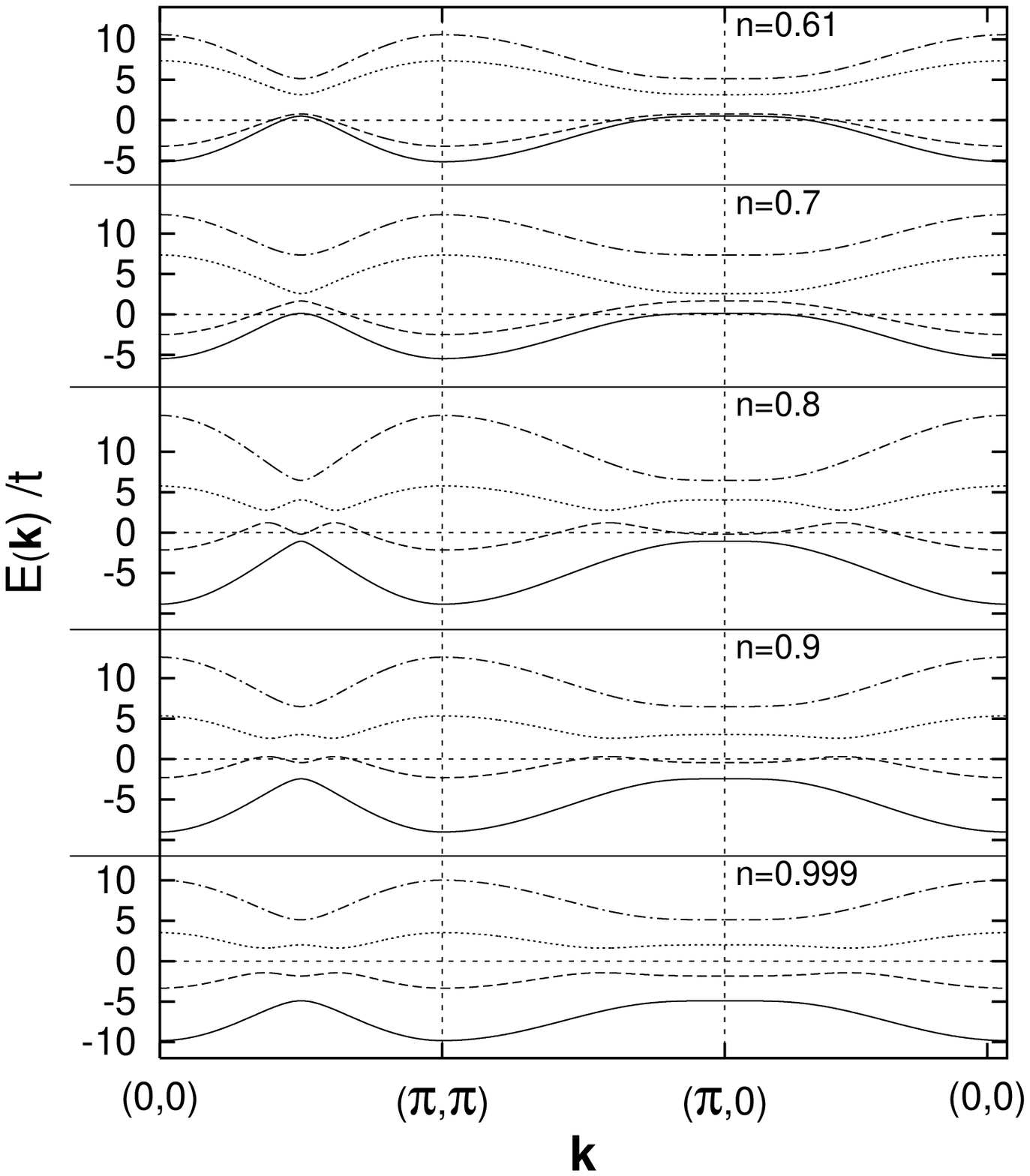}}
  \caption{Sublattice density of states and band structure at $kT=0.4t$ and with
    $V=U/5$; the thin vertical line at $E=0$ denotes the position of the
    Fermi level $E_F$.}
  \label{fig:21}
\end{figure}
Two peaks are due to the central gap and two peaks are reminiscent of the van
Hove singularity in each Hubbard band \cite{Avella:97b} (cf.\
Fig.~\ref{fig:21}).
\begin{figure}[tbp]
  \centering
  \includegraphics*[width=4.5cm,angle=270]{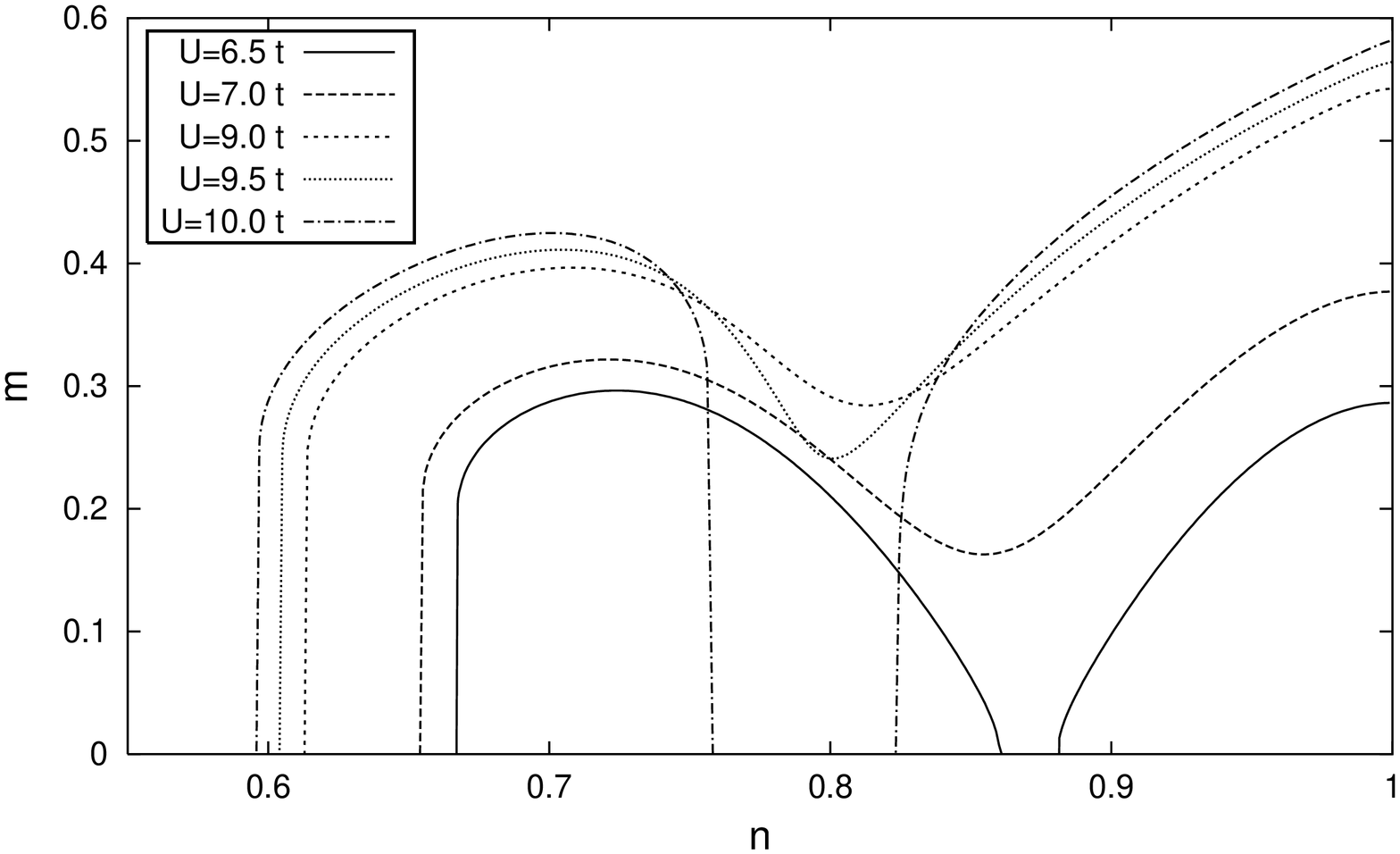}
  \caption{Sublattice magnetization for different values of the Coulomb interaction at
    $kT=0.4t$ with $V=U/5$.}
  \label{fig:22}
\end{figure}
In Fig.~\ref{fig:22} we plotted the sublattice magnetization as a function of
the particle density for various values of the Coulomb interaction $U$ showing
the separation of the antiferromagnetic phase by a paramagnetic region for
high as well as for low values of $U$.

\begin{figure}[tbp]
  \centering
  \includegraphics*[width=4.5cm,angle=270]{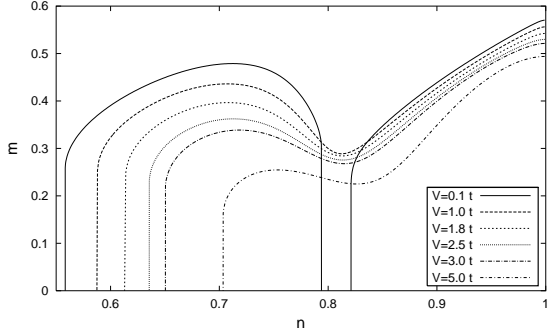}
  \caption{Sublattice magnetization as function of the particle density for different
    values of the inter-site Coulomb interaction $V$, with $U=9t$ and $kT=0.4t$}
  \label{fig:23}
\end{figure}

The inter-site Coulomb interaction $V$ considerably reduces the magnetization
and the stability of the antiferromagnetic state, i.e., its extension in $n$
as can be seen from Fig.~\ref{fig:23}. The reduction of the magnetization also
suppresses the separation of the two parts of the antiferromagnetic states.
However, the suppression of antiferromagnetism by the inter-site Coulomb
interaction is much smaller than the one found in the treatment of the simple
Hubbard model.

\section{Conclusions}
\label{sec:6}

Among the variety of analytical methods, developed in the last decades to deal
with strongly correlated electron systems, the \emph{COM} has been rather
successful in describing the properties of many correlated systems
\cite{Mancini:00b,Ins,Mancini:95,Avella:98,Matsumoto,Ishihara:94}. To add
another piece to the puzzle constituted by the phase diagram of the Hubbard
model, we here investigated the antiferromagnetic phase characterized by a
staggered magnetization. A fully self-consistent treatment, respecting the
symmetry constrains emerging from the Pauli principle, has been presented for
the Hubbard model in two and three dimensions and the two-dimensional extended
Hubbard model.

The antiferromagnetic phases of the three systems, where compared to the
corresponding paramagnetic phases, lead to a phase transition of the second
order. In all cases the antiferromagnetic states have lower free energy.
Furthermore, the antiferromagnetic states of all the models show the following
properties:
\renewcommand{\theenumi}{\alph{enumi}}
\renewcommand{\labelenumi}{(\theenumi)}
\begin{enumerate}
\item the presence of three kinds of transitions (Mott-Hubbard, Mott-Heisenberg and
Heisenberg) at half-filling in the plane $T$-$U$;
\item the existence of two components in the antiferromagnetic gap (one due to
the antiferromagnetic correlations and another coming from the Mott-Hubbard
mechanism);
\item a finite critical value of the Coulomb interaction for the Mott-Hubbard and
Mott-Heisenberg transitions;
\item the antiferromagnetic phase is stable only in a very narrow region around half
filling, showing a strong reduction of the N{\'e}el temperature with doping;
\item a metal-insulator transition, driven by the temperature, takes
place within the antiferromagnetic phase; at half filling and higher
temperatures this transition coincides with the paramagnetic-antiferromagnetic
one;
\item away from half-filling a metal-insulator transition driven by the doping is observed.
This transition has the following properties: at zero temperature it is
discontinuous and connects an antiferromagnetic metal and an antiferromagnetic
insulator; at finite temperature it connects the antiferromagnetic metal to an
antiferromagnetic state of semiconductor type.
\end{enumerate}
All these properties emerge from very strong correlations and are not usually
found by approximations of mean-field type. They could be explained by
analyzing the electronic density of states and the energy spectra.

Finally, we want to point out that the thermodynamical study of the Hubbard
model within the framework of the two-pole approximation by means of the
\emph{COM} is far from being completed. Besides the normal and the
antiferromagnetic phase, a detailed analysis of the superconducting
\cite{Super}, the ferromagnetic \cite{Avella:00a}, the charge ordered and of
other phases with more complex magnetic and charge ordering has to be
completed or undertaken and these analysis should finally combine to give the
complete phase diagram of the Hubbard model under this approach.

\acknowledgments

This work has been partially supported by the {\em Studienstiftung des
deutschen Volkes}. One of the authors (R.M.) wants to thank Prof.\ F.\ Mancini
and the {\em Dipartimento di Fisica ``E.R. Caianiello''} at the University of
Salerno for the hospitality and the motivating atmosphere during the period he
stayed at the Institute. Furthermore, we want to thank Prof.\ W.\ Nolting and
Dr.\ V.\ Turkowski for helpful discussions.

%\bibliographystyle{prsty}
%\bibliography{biblio}

\appendix
\section{Calculation of the Green's Functions for the
  \lowercase{$\mathbf t$--$\mathbf t^{\prime}$}--$\mathbf U$ Model}
\label{sec:A.1}

\begin{figure}[tbp]
  \centering
  \includegraphics*[width=6.0cm,angle=0]{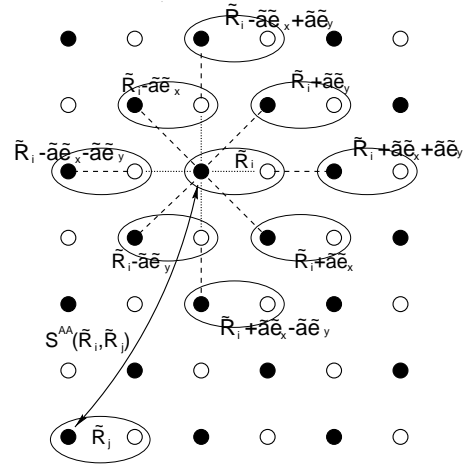}
  \vskip 5mm
  \caption{The definition of the translational invariant Green's function $S^{AA}$.}
  \label{fig:A.1.1}
\end{figure}

On the magnetic lattice defined in Fig.~\ref{fig:2} we can define the
translational invariant Green's functions $S^{AA}( \widetilde{{\mathbf
R}}_i,\widetilde{{\mathbf R}}_j,t)$, $S^{AB}( \widetilde{{\mathbf
R}}_i,\widetilde{{\mathbf R}}_j,t)$, $S^{BA} ( \widetilde{{\mathbf
R}}_i,\widetilde{{\mathbf R}}_j,t)$ and $S^{BB}( \widetilde{{\mathbf
R}}_i,\widetilde{{\mathbf R}}_j,t)$ connecting two points $\widetilde{{\mathbf
R}}_i$ and $\widetilde{{\mathbf R}}_j$ of the magnetic lattice. Their
definition and the corresponding equations of motion are illustrated in
Fig.~\ref{fig:A.1.1} for the translational invariant Green's function $S^{AA}$.
By means of the Fourier transform on the magnetic lattice
\end{multicols} \widetext
\begin{equation}
  \label{eq:a1.2}
  S^{XY}( \widetilde{{\mathbf R}}_i,\widetilde{{\mathbf R}}_j,t)
  =S^{XY}( \widetilde{{\mathbf R}}_i-\widetilde{{\mathbf R}}_j,t)
  =\frac i{2\pi }\Bigl( \frac{\widetilde{a}^2}{\left
      ( 2\pi \right) ^2}\Bigr)
  ^2\int {\rm d}\omega \, e^{-i\omega t}\int_{\Omega _{\widetilde{B}}}{\rm
  d}^2\widetilde{k}\; e^{i\widetilde{{\mathbf k}}{\mathbf \cdot }( \widetilde{{\mathbf R}}_i-
      \widetilde{{\mathbf R}}_j) }S^{XY}( \widetilde{{\mathbf k}}{\mathbf ,
      }\omega ) \quad ,\quad X,Y\in \left\{ {\mathbf A},{\mathbf B}\right\}
\end{equation}
the equations of motion for the translational invariant Green's functions take
the form
\begin{equation}
  \label{eq:a1.3}
  \begin{split}
    \omega S^{AA}( \widetilde{{\mathbf k}},\omega )  &=\bigl[
      I^{\left( n\right) }+I^{\left( m\right) }\bigr] +\Bigl( \bigl[ \varepsilon
        ^{(1)}+\varepsilon ^{(3)}\bigr] +\widetilde{\beta }( \widetilde{{\mathbf k}}
      ) \bigl[ \varepsilon ^{(5)}+\varepsilon ^{(6)}\bigr] \Bigr) S^{AA}(
      \widetilde{{\mathbf k}},\omega ) +\bigl( \widetilde{\alpha }(
        \widetilde{{\mathbf k}}) \bigr) ^{*}\bigl[ \varepsilon ^{(2)}+\varepsilon
      ^{(4)}\bigr] S^{BA}( \widetilde{{\mathbf k}},\omega )  \\
    \omega S^{AB}( \widetilde{{\mathbf k}},\omega )  &=\Bigl( \bigl[
        \varepsilon ^{(1)}+\varepsilon ^{(3)}\bigr] +\widetilde{\beta }( \widetilde{
          {\mathbf k}}) \bigl[ \varepsilon ^{(5)}+\varepsilon ^{(6)}\bigr] \Bigr)
    S^{AB}( \widetilde{{\mathbf k}},\omega ) +\bigl( \widetilde{\alpha
        }( \widetilde{{\mathbf k}}) \bigr) ^{*}\bigl[ \varepsilon
      ^{(2)}+\varepsilon ^{(4)}\bigr] S^{BB}( \widetilde{{\mathbf k}},\omega )
    \\
    \omega S^{BA}( \widetilde{{\mathbf k}},\omega )  &=\Bigl( \bigl[
        \varepsilon ^{(1)}-\varepsilon ^{(3)}\bigr] +\widetilde{\beta }( \widetilde{
          {\mathbf k}}) \bigl[ \varepsilon ^{(5)}-\varepsilon ^{(6)}\bigr] \Bigr)
    S^{BA}( \widetilde{{\mathbf k}},\omega ) +\widetilde{\alpha }
    ( \widetilde{{\mathbf k}}) \bigl[ \varepsilon ^{(2)}-\varepsilon
      ^{(4)}\bigr] S^{AA}( \widetilde{{\mathbf k}},\omega )  \\
    \omega S^{BB}( \widetilde{{\mathbf k}},\omega )  &=\bigl[
      I^{\left( n\right) }-I^{\left( m\right) }\bigr] +\Bigl( \bigl[ \varepsilon
        ^{(1)}-\varepsilon ^{(3)}\bigr] +\widetilde{\beta }( \widetilde{{\mathbf k}}
      ) \bigl[ \varepsilon ^{(5)}-\varepsilon ^{(6)}\bigr] \Bigr) S^{BB}(
      \widetilde{{\mathbf k}},\omega ) +\widetilde{\alpha }( \widetilde{
        {\mathbf k}}) \bigl[ \varepsilon ^{(2)}-\varepsilon ^{(4)}\bigr] S^{AB}(
      \widetilde{{\mathbf k}},\omega )
  \end{split}
\end{equation}
\begin{multicols}{2} \narrowtext \noindent where we used the projections on the magnetic lattice
$\widetilde{\alpha } ( \widetilde{{\mathbf k}}) =\frac 14 [
1+e^{i\widetilde{k}_x\widetilde{a}}+e^{i\widetilde{k}_y\widetilde{a}
  }+e^{i( \widetilde{k}_x\widetilde{a}+\widetilde{k}_y\widetilde{a}
  ) }]$ and $\widetilde{\beta }
( \widetilde{{\mathbf k}}) =\frac 13( 4| \widetilde{\alpha
  }( \widetilde{{\mathbf k}}) |^2-1)$ from Eq.~\eqref{eq:3.24}.
The expressions~\eqref{eq:3.21} and \eqref{eq:3.22} are obtained from
Eqs.~\eqref{eq:a1.3} by lengthy but straightforward algebraic manipulations.
\section{Explicit Expressions for the Green's Functions of the Antiferromagnetic
  Equilibrium State}
\label{sec:A.2} For $ X,Y\in \left\{ {\mathbf A},{\mathbf B}\right\}$ the
coefficients $A^{XY}$, $B^{XY}$, $C^{XY}$ occurring in the
expression~\eqref{eq:3.21} of the Green's functions take the explicit form:
\begin{align*}
  A^{AA}&=-{\cal E}^{+}_1-{\cal E}^{+}_2{\cal E}^{-}_1\frac 1{{\cal E}^{+}_2}-\widetilde{\beta }(
    \widetilde{{\mathbf k}}) \Bigl[ {\cal E}^{+}_3+{\cal E}^{+}_2{\cal E}^{-}_3\frac
    1{{\cal E}^{+}_2}\Bigr]  \\
  B^{AA}&={\cal E}^{+}_2{\cal E}^{-}_1\frac 1{{\cal E}^{+}_2}{\cal E}^{+}_1+\widetilde{\beta }(
    \widetilde{{\mathbf k}}) \Bigl[ {\cal E}^{+}_2{\cal E}^{-}_3\frac
    1{{\cal E}^{+}_2}{\cal E}^{+}_1\\
    &\qquad+{\cal E}^{+}_2{\cal E}^{-}_1\frac 1{{\cal E}^{+}_2}{\cal E}^{+}_3\Bigr]
  +\bigl(\widetilde{\beta }( \widetilde{{\mathbf k}}) \bigr)
  ^2{\cal E}^{+}_2{\cal E}^{-}_3\frac 1{{\cal E}^{+}_2}{\cal E}^{+}_3\\
  &\qquad-\lvert \widetilde{\alpha }(
      \widetilde{{\mathbf k}}) \rvert ^2{\cal E}^{+}_2{\cal E}^{-}_2 \\
  C^{AA}&=I^{\left( n\right) }+I^{\left( m\right) } \\
  D^{AA}&=-\Bigl[ {\cal E}^{+}_2{\cal E}^{-}_1+\widetilde{\beta }( \widetilde{{\mathbf k}
        }) {\cal E}^{+}_2{\cal E}^{-}_3\Bigr] \frac 1{{\cal E}^{+}_2}\Bigl[ I^{\left( n\right)
      }+I^{\left( m\right) }\Bigr]
\end{align*}
\begin{align*}
  A^{BB}&=-{\cal E}^{-}_1-{\cal E}^{-}_2{\cal E}^{+}_1\frac 1{{\cal E}^{-}_2}-\widetilde{\beta }(
    \widetilde{{\mathbf k}}) \Bigl[ {\cal E}^{-}_3+{\cal E}^{-}_2{\cal E}^{+}_3\frac
    1{{\cal E}^{-}_2}\Bigr]  \\
  B^{BB}&={\cal E}^{-}_2{\cal E}^{+}_1\frac 1{{\cal E}^{-}_2}{\cal E}^{-}_1+\widetilde{\beta }(
    \widetilde{{\mathbf k}}) \Bigl[ {\cal E}^{-}_2{\cal E}^{+}_3\frac
    1{{\cal E}^{-}_2}{\cal E}^{-}_1\\
    &\qquad+{\cal E}^{-}_2{\cal E}^{+}_1\frac 1{{\cal E}^{-}_2}{\cal E}^{-}_3\Bigr] +\bigl(
    \widetilde{\beta }( \widetilde{{\mathbf k}}) \bigr)
  ^2{\cal E}^{-}_2{\cal E}^{+}_3\frac 1{{\cal E}^{-}_2}{\cal E}^{-}_3\\
  &\qquad-\lvert \widetilde{\alpha }(
      \widetilde{{\mathbf k}}) \rvert ^2{\cal E}^{-}_2{\cal E}^{+}_2 \\
  C^{BB}&=I^{\left( n\right) }-I^{\left( m\right) } \\
  D^{BB}&=-\Bigl[ {\cal E}^{-}_2{\cal E}^{+}_1+\widetilde{\beta }( \widetilde{{\mathbf k}
    }) {\cal E}^{-}_2{\cal E}^{+}_3\Bigr] \frac 1{{\cal E}^{-}_2}\Bigl[ I^{\left( n\right)
    }-I^{\left( m\right) }\Bigr]
\end{align*}
\begin{align*}
  A^{AB}&=A^{AA} \\
  B^{AB}&=B^{AA}\\
  C^{AB}&=0 \\
  D^{AB}&=\bigl( \widetilde{\alpha }( \widetilde{\mathbf{k}})
  \bigr) ^{*}{\cal E}^{+}_2\Bigl[ I^{\left( n\right) }-I^{\left( m\right) }\Bigr]
\end{align*}
\begin{align*}
  A^{BA}&=A^{BB}\\
  B^{BA}&=B^{BB}\\
  C^{BA}&=0 \\
  D^{BA}&=\widetilde{\alpha }( \widetilde{\mathbf{k}})
  {\cal E}^{-}_2\Bigl[ I^{\left( n\right) }+I^{\left( m\right) }\Bigr]
  \, .\\[-0.5ex]
  \hfill\label{eq:3.22}\tag{\theequation}\addtocounter{equation}{1}
\end{align*}
To abbreviate the notation we used the following definitions
\begin{equation}
  \label{eq:3.23}
  \begin{array}{lll}
    {\cal E}^{+}_1=\varepsilon ^{(1)}+\varepsilon ^{(3)} & \quad , \quad & {\cal E}^{-}_1=\varepsilon
    ^{(1)}-\varepsilon ^{(3)} \\
    {\cal E}^{+}_2=\varepsilon ^{(2)}+\varepsilon ^{(4)} & \quad ,  \quad & {\cal E}^{-}_2=\varepsilon
    ^{(2)}-\varepsilon ^{(4)} \\
    {\cal E}^{+}_3=\varepsilon ^{(5)}+\varepsilon ^{(6)} & \quad ,  \quad & {\cal E}^{-}_3=\varepsilon
    ^{(5)}-\varepsilon ^{(6)} \, .
  \end{array}
\end{equation}
Furthermore, we used the convention that $\cos \left( {\mathbf Q}\cdot
{\mathbf R}_i\right)$ takes positive values on the sublattice $\mathbf A$, and
we notice that the projections on the nearest neighbors and on the next- and
next-next-nearest neighbors according to the spherical approximation are
  given on the magnetic lattice by the expressions
\begin{equation}
  \label{eq:3.24}
  \begin{split}
    \widetilde{\alpha }(\widetilde{\mathbf k}) =& \frac 14\left( 1+e^{i\widetilde{k}_x
        \widetilde{a}}+e^{i\widetilde{k}_y\widetilde{a}}+e^{i(\widetilde{k}_x
        \widetilde{a}+\widetilde{k}_y \widetilde{a}) }\right)  \\
    \widetilde{\beta }( \widetilde{\mathbf k}) =& \frac 16\left( e^{i\widetilde{k}_x
        \widetilde{a}}+e^{-i\widetilde{k}_x\widetilde{a}}+e^{i\widetilde{k}_y
        \widetilde{a}}+e^{-i\widetilde{k}_y
        \widetilde{a}}\right)\\
    & + \frac 1{12}\left( e^{i( \widetilde{k}_x\widetilde{a}+\widetilde{k}_y
          \widetilde{a}) }+e^{-i( \widetilde{k}_x\widetilde{a}
        +\widetilde{k}_y\widetilde{a})
        }+e^{i( \widetilde{k}_x\widetilde{a}-\widetilde{k}_y\widetilde{a}) }\right .\\
    &\qquad\qquad\left . +e^{-i( \widetilde{k}_x
          \widetilde{a}-\widetilde{k}_y\widetilde{a}) }\right) \, .
  \end{split}
\end{equation}

\end{multicols}

\end{document}